\documentclass[english,a4paper,11pt]{article}
\usepackage[english]{babel}
\usepackage{graphicx,epsfig,epic} % standard LaTeX graphics tool
\usepackage{pifont}      % placeholder for figures

\usepackage{pict2e}
\usepackage{amssymb}

\font\tenbb=msbm10 at 12pt
\def\cC{\hbox{\tenbb C}}
\def\rR{\hbox{\tenbb R}}
\def\nN{\hbox{\tenbb N}}

\def\cal{\mathcal}
\def\esp{\vskip .6cm}

\newtheorem{thm}{Theorem}

\newtheorem{lem}{Lemma}

\newtheorem{rem}{Remark}

\newsavebox{\fmbox}

\begin{document}

%%%% Article title to be placed here
\title{Fluctuating Paths of Least Time, Schr\"{o}dinger Equation, Van der Waals torque and Casimir Effect Mechanism}

\author{%%%% Author details
F. Ben Adda\footnote{Community College of Qatar, Department of math and Sciences, Doha, Qatar}}

\maketitle

%%%% Abstract text to be placed here %%%%%%%%%%%%
\begin{abstract}
The use of an infinity of fluctuating paths of least time that are compatible with the quantum mechanics indeterminacy provides a new interpretation in geometrical optic of the interference pattern of Young's double slit experiment, which suggests that the wave behavior of matter and radiation is dictated by the space-time geodesics. Moreover, the association of a wave function to each path of least time as a probability amplitude together with an uncertainty for momentum and position allows to derive the Schr\"{o}dinger's equation starting from the geodesic's characteristics. A new insight is obtained regarding the Van Der Waals torque as well as Casimir attraction/repulsion mechanism.
\end{abstract}
%%%%%%%%%%%%%%%%%%%%%%%%%%%


%{\small Keywords}: {\small .}

\esp
{\small\tableofcontents}
%%%%%%%%%% Insert the texts which can accomdate on firstpage in the tag "fmtext" %%%%%

\section{Introduction}
%%%% Insert A head here
The most challenging problem of modern physics is to find a coherent geometrical representation of the space-time at all scales. The real geometry and architecture of the space-time of our universe and the constraints that it imposes to its content are still unknown (specially for the space-time at the quantum scale (\cite{1}) despite the fact that local descriptions of the space-time geometry provide a suitable understanding of the behavior of several physical systems on it to a certain limit. However, without knowing the real constraints that the space-time geometry and characteristics impose to its content, we may not be able to acquire a precise understanding of certain physical systems behavior. At the quantum scale Heisenberg, Born and Jordan have developed a matrix formalism of quantum mechanics in 1925 (\cite{18},\cite{19},\cite{20}) in which matrix was used as a physical property of elementary particles. In 1926 Schr\"{o}dinger introduced another approach of quantum mechanics(\cite{21,22,23,24}), where the physical properties of elementary particles were described by a wave function solution of a differential equation. The quantum formalism found its development on an infinitely dimensional Hilbert space of functions where the quantum description of a point particle was completely unusual for a trajectory basic understanding since it prohibits the simultaneous determination of position and momentum with accurate measurement, which suggests that the classical trajectory of a particle has no physical meaning in the quantum mechanic's interpretation. Nevertheless, in 1948 Feynman introduced a different approach of quantum mechanics using the path integral (\cite{9}) as a sum over all possible trajectories to calculate a total quantum amplitude compatible with the Schr\"{o}dinger's equation. Feynman's path integral conveys in its formalism a connection between Lagrangian and quantum mechanics. Thus, despite the correct multiple predictions of the Feynman's path integral in physics, a lack of mathematical rigor exists for the geometrical interpretation of the integration over all possible paths.

In this case study we use the essence of the Feynman's path integral in a different way that involves the medium characteristics together with Heisenberg uncertainty relation to provide a fully equivalent meaning compatible with the Schr\"{o}dinger's equation in order to elaborate a new insight for a better understanding of the Van Der Waals torque (\cite{7},\cite{10}) as well as the Casimir effect (\cite{5}) at the quantum scale. Many thoughts about the origin and mechanism of the phenomenon of Casimir effect and Van Der Waals torque are still under intense investigation experimentally and theoretically for their utility in nano-mechanics, nano-technology (\cite{6}). A controversy exists about the origin of this phenomenon (\cite{5},\cite{11}, \cite{12}, \cite{13}, \cite{16}), and this case study provides a new insight that might help to forge a clear understanding.

The plan of this paper is as follow: in section 2, a brief graphic introduction is given about infinity of paths of least time. In section 3, reproduction of interference pattern of Young's double-slit experiment using paths of least time and interpretation is given in geometrical optic. In section 4, a mathematical formulation of paths of least time is presented in 2D, 3D, as well as properties. In section 5, action and wave function are associated to each geodesic following the essence of Feynman's path integral, and formulation of an uncertainty relation. In section 6, the derivation of Schr\"{o}dinger Equation is obtained starting from characteristics of the infinity of paths of least time in the complex plan. In section 7, approach to the zero point energy, Van Der Waals torque and Casimir effect via Schr\"{o}dinger Equation are elaborated. As an outcome a new insight that concerns causality and interpretation of Van Der Waals torque as well as mechanism of attractive/repulsive Casimir force is figured out. In section 8, conclusion.
\section{Graphical introduction of geodesics in an homogeneous, isotropic and expanding space-time}

\begin{figure}[!h]
\begin{minipage}[t]{6cm}
\centering
\includegraphics[width=6cm]{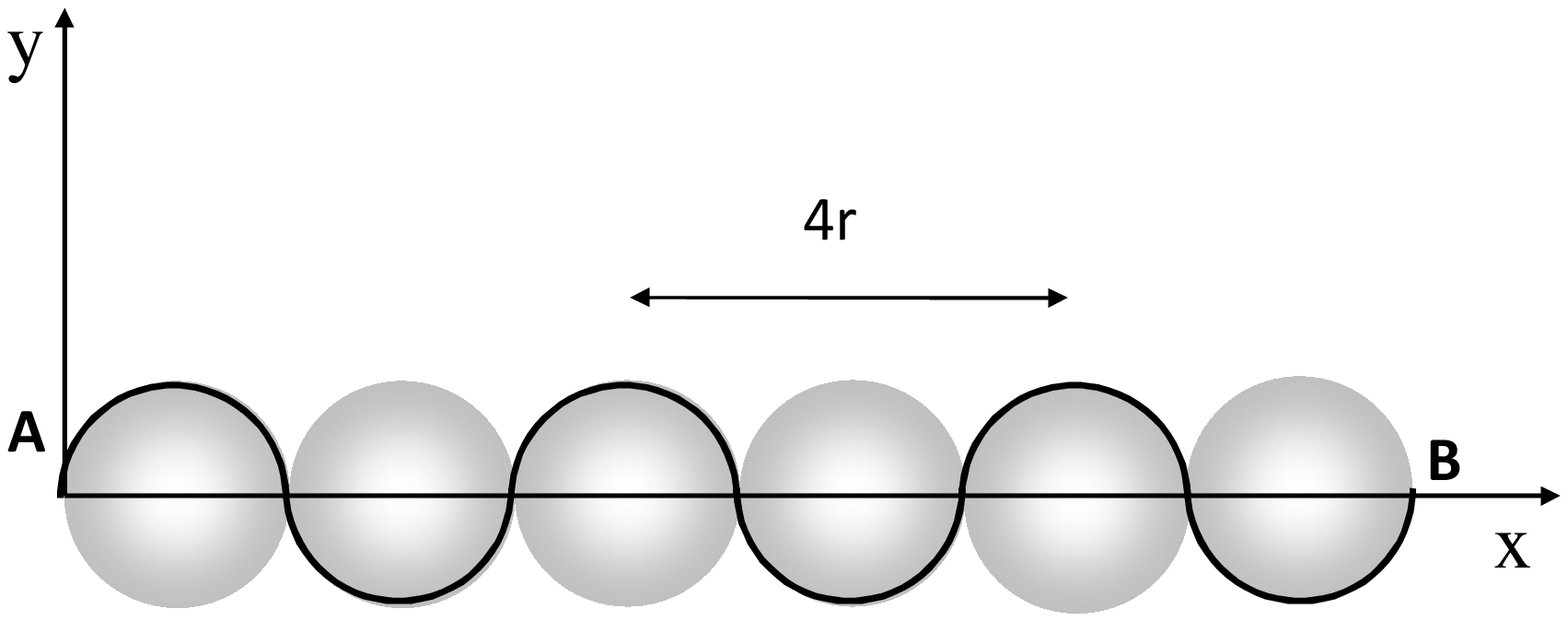}
\caption{\footnotesize Illustration of one geodesic in 2D for $N=6$ (alignment of 6 spheres) and arc radius $r=5\ mm$ between aligned antipodal points from A to B.}\label{Fig.0}
\end{minipage}
\hspace*{\fill}
\begin{minipage}[t]{6cm}
\centering
\includegraphics[width=6cm]{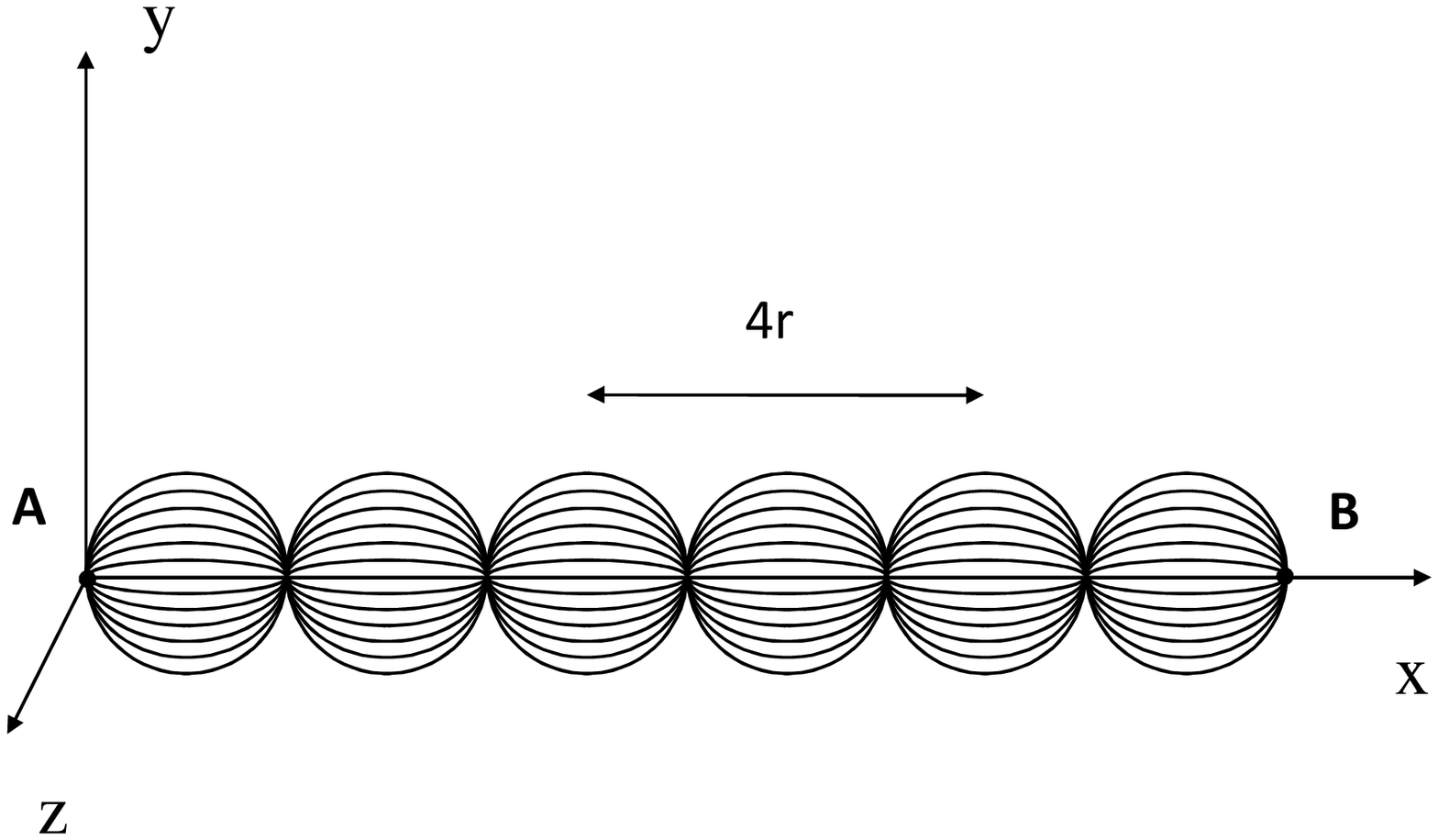}
\caption{\footnotesize Illustration of infinity of geodesics in 3D with identical length for $N=6$ and arc radius $r=5\ mm$ between aligned antipodal points from A to B. }\label{Fig.00}
\end{minipage}
\end{figure}

Observation at large scale revealed confirmed characteristics of the space-time: it is an homogeneous, isotropic, expanding and non homogeneous locally space-time. A case study (\cite {3}) based on a simulation of an homogeneous, isotropic and expanding space time that expands via expansion of its basic elements led to approach an interesting property of the space-time geodesics at small and large scale;
the geodesics (or paths of least time) in an homogeneous, isotropic and expanding space-time that expands via expansion of its basic elements cannot be straight line geodesics, rather it is an infinity of fluctuating geodesics with non null curvature illustrated in (Fig.\ref{Fig.00}). These geodesics allow to approach the quantum scale.

\begin{figure}[!h]
\begin{minipage}[t]{6cm}
\centering
\includegraphics[width=6cm]{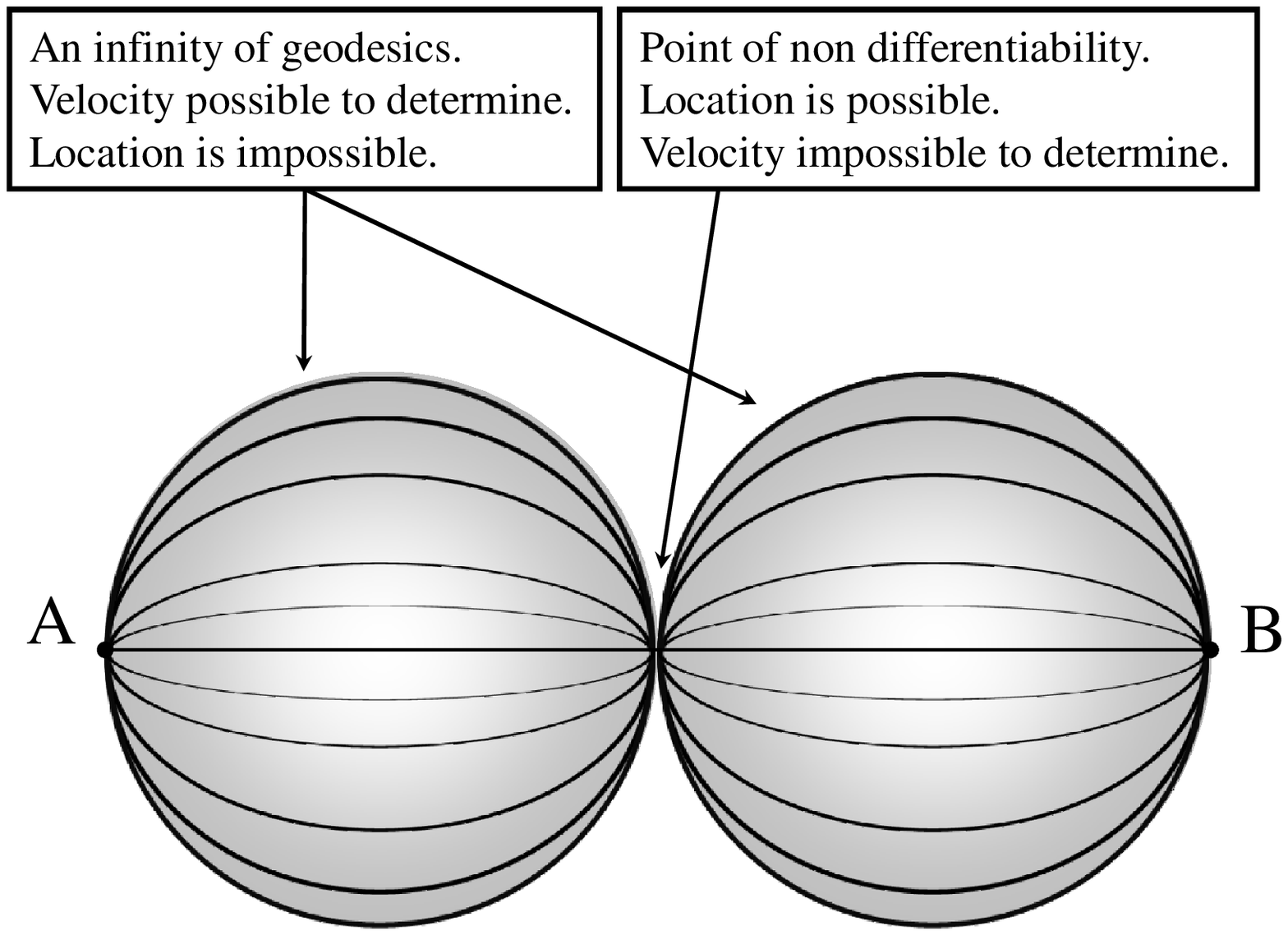}
\caption{\footnotesize Consistency with the indeterminacy of quantum mechanics. There exists an infinity of geodesics with non null curvature between two distant locations consistent with the indeterminacy of quantum mechanics}\label{Fig.1}
\end{minipage}
\hspace*{\fill}
\begin{minipage}[t]{6cm}
\centering
\includegraphics[width=6cm]{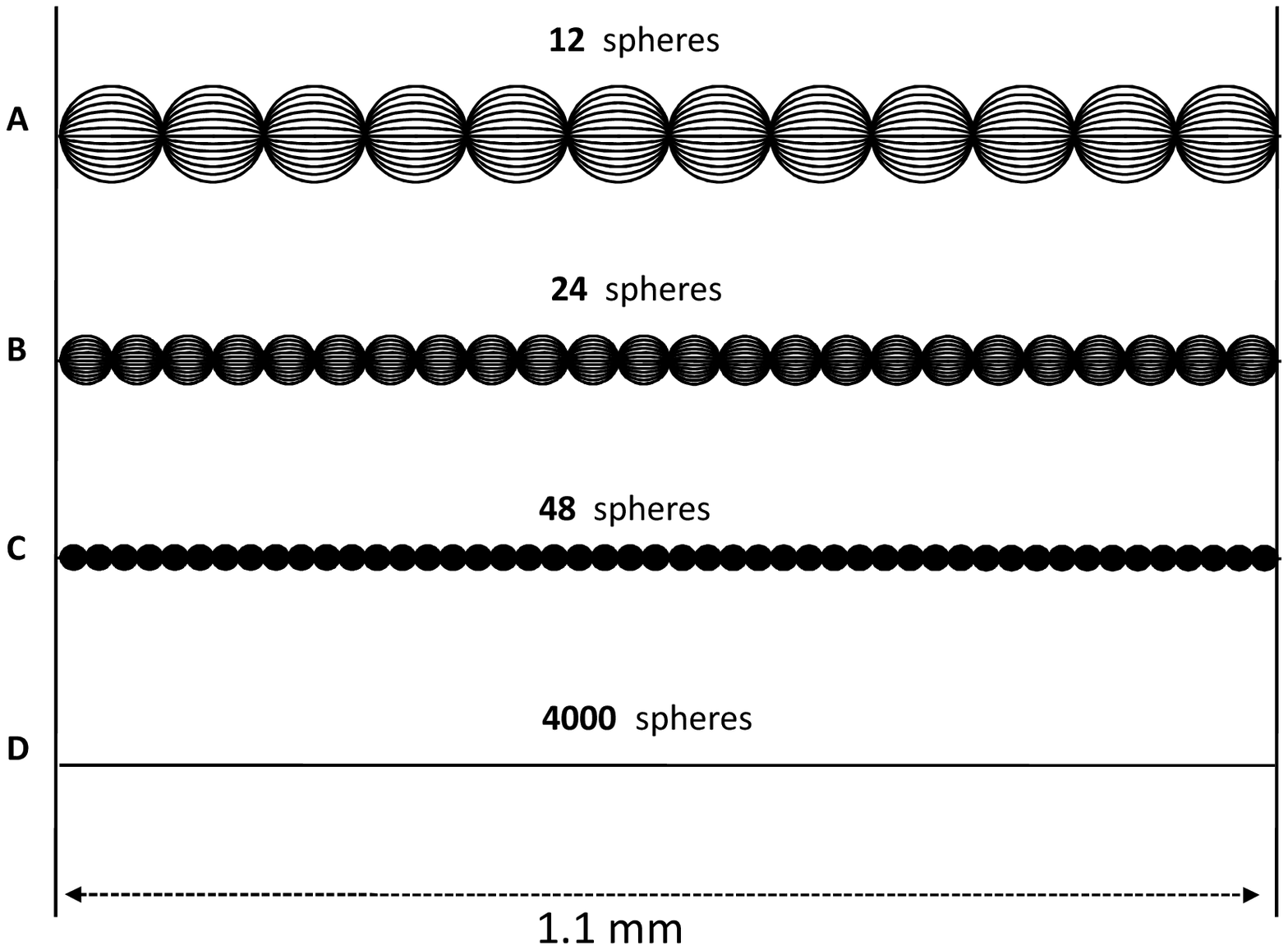}
\caption{\footnotesize Alinement of spheres in $1.1 mm$. A) 12 spheres of radius $r_A\simeq0,0458333mm$, B) 24 spheres of radius $r_B\simeq0,0229166mm$, C) 48 spheres of radius $r_C\simeq0,0114583mm$, D) 4000 spheres of radius $r_D\simeq0,0001375mm$. }\label{Fig.2}
\end{minipage}
\end{figure}

Indeed, the infinity of fluctuating geodesics illustrated in (Fig.\ref{Fig.00}) between two distant locations have the same length, and are continuous, differentiable, with a finite number of points of non differentiability (antipodal points), where localization on it is impossible between any two antipodal points due to the existence of an infinity of circle arcs with equal length that prohibit localization, which makes these geodesics consistent with the indeterminacy of quantum mechanics (Fig.\ref{Fig.1}).
Moreover, it is known that within 1.1 mm there is an average of 2000 light wave lengths that correspond to 4000 aligned spheres of radius $r_0=0.0001375 mm$ using the geodesics illustrated in (Fig.\ref{Fig.00}). The aligned 4000 spheres of radius $r_0$ within 1.1 mm appear at the macroscopic observation as a straight segment (the fluctuations are undiscernible), which fits the local description of a space-time with null curvature (see Fig.\ref{Fig.2}), meanwhile a microscopic view of them reveals the existence of infinity of transverse fluctuating geodesics with points of non differentiability. This characteristics might reveal new details that concerns Young's double-slit experiment.

Indeed,
It is known that the double-slit experiment is a corner stone in quantum mechanics and plays a crucial role in our comprehension of the quantum world mystery. It is also known that geometrical optic failed drastically to explain it. Thus, the Young's double-slit experiment (\cite{14},\cite{15}) conveys a fundamental mystery in physics: how can light be emitted and absorbed as corpuscular, and undergo interference pattern between source and detector screen as a wave? The adoption of duality allows to circumvent this question and provides flexible approach in our use of the corpuscular character or wave nature of matter and radiation following the needs. Because of this mystery, testing the infinity of fluctuating geodesics (Fig.\ref{Fig.00}) with the interference pattern seems to be the first step and might provide a new interpretation for the phenomenon in geometrical optic. Meanwhile proving that these infinity of geodesics convey a fully equivalent formalism compatible with the Schr\"{o}dinger's equation will allow to approach energy fluctuation, zero point energy, Van der Waals torque as well as the Casimir effect from a new perspective.

\section{Infinity of fluctuating paths of least time and interference pattern in Young's double-slit experiment}\label{IPGEO}

\begin{figure}[!h]
\begin{minipage}[t]{6cm}
\centering
\includegraphics[width=6cm]{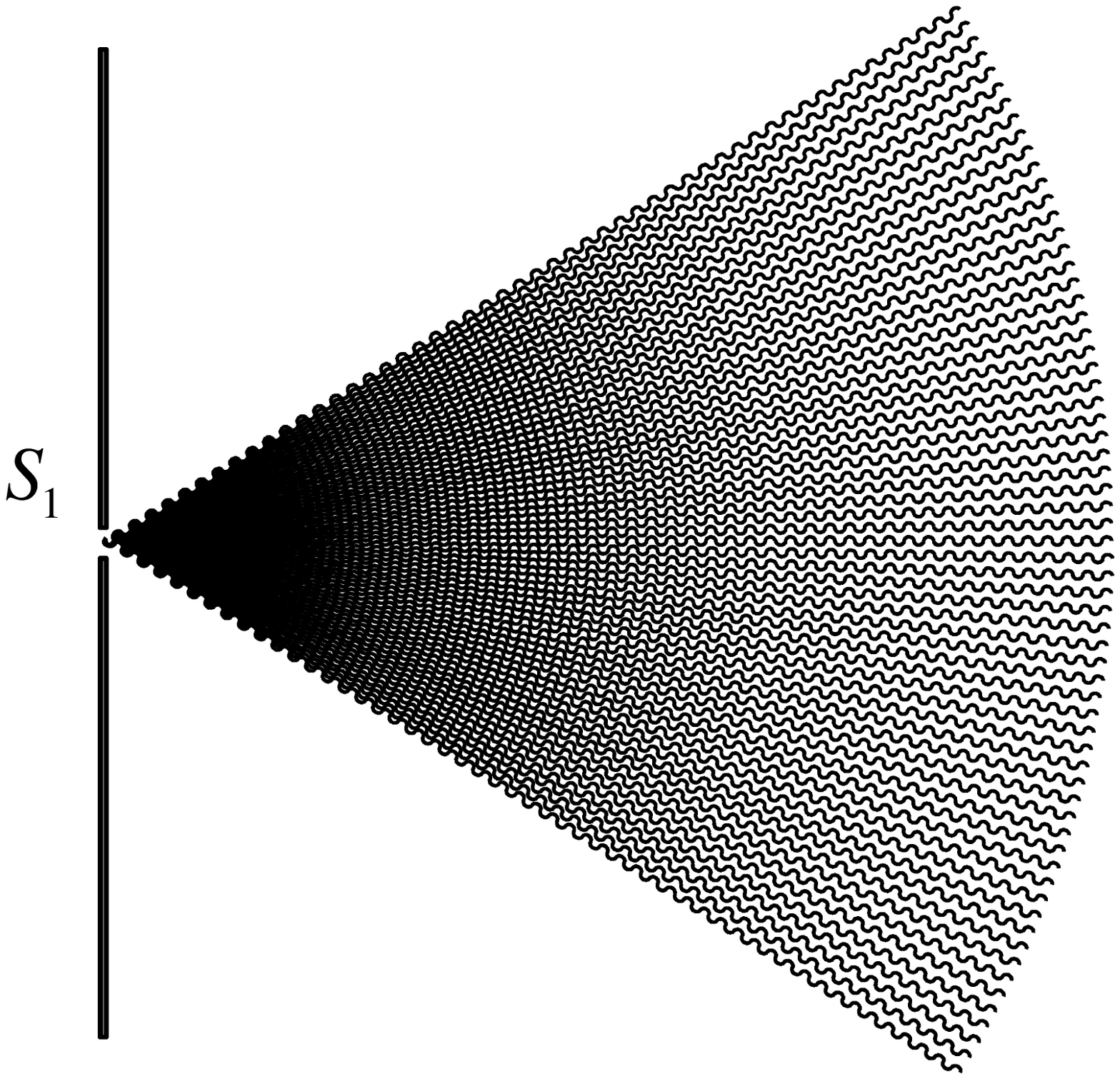}
\caption{\footnotesize Geodesics diffraction: Illustration of 65 geodesics in phase, using the geodesic given by (Fig.\ref{Fig.0}) for $r_0=0.25\ mm$, N=102, that cover an angle of $64^\circ$ from a narrow slit, to represent all possible geodesics that diffract from the slit. The angle between two consecutive geodesic in phase is $1^\circ$ .}\label{Fig.3}
\end{minipage}
\hspace*{\fill}
\begin{minipage}[t]{6cm}
\centering
\includegraphics[width=6cm]{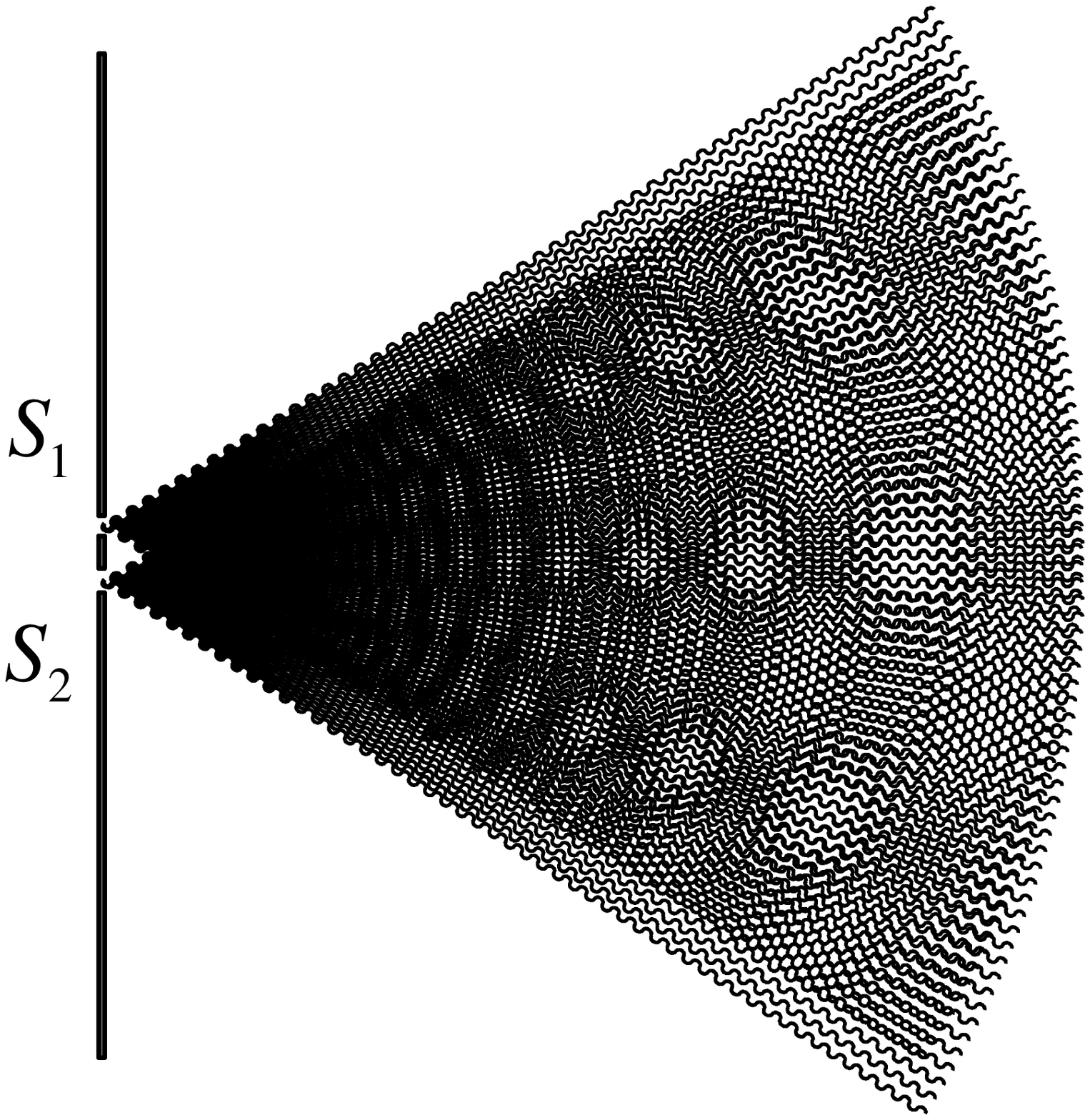}
\caption{\footnotesize The superposition of family of geodesics of radius $r_0=0.25mm$ from the $S_1$ with the family of geodesics with the same radius from the slits $S_2$ generate fringes or bands in the geodesics intersection region. }\label{Fig.4}
\end{minipage}
\end{figure}

To simulate all possible paths of least time in phase that flare out from a given slit $S_1$ and cover a chosen angle of $\theta=64^\circ$ beyond the slit $S_1$ as illustrated in (Fig.\ref{Fig.3}), we will use only one plan geodesic from the infinity of geodesics illustrated in (Fig.\ref{Fig.00}). The chosen angle can be covered by 65 copies of the same plan geodesic represented in (Fig.\ref{Fig.0}) for $r_0=0.25\ mm$, N=102 (ie: on 102 aligned spheres) where the angle between two successive plan geodesics is one degree using a rotation in the plan about $S_1$ (interpretation of the diffraction using the fluctuating geodesics can be found \cite{2} for more details).

\begin{figure}[!h]
\begin{minipage}[t]{6cm}
\centering
\includegraphics[width=6cm]{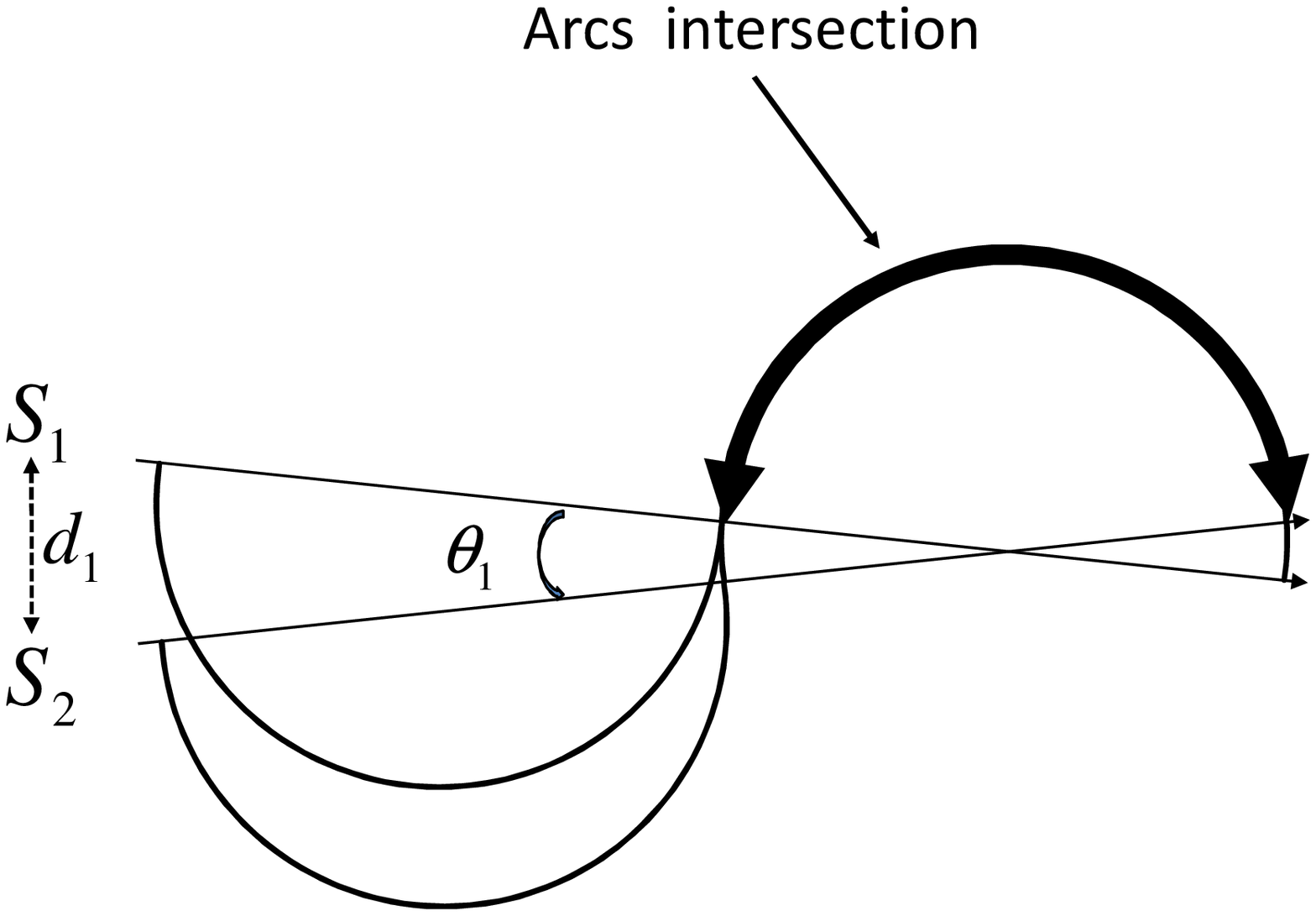}
\caption{\footnotesize Zoom on arcs intersection that produce interference pattern in the geodesics intersection region between slits screen and detector screen. Large geodesics arcs intersection for reduced slits distance $d_1=d(S_1,S_2)$.}\label{Fig.7}
\end{minipage}
\hspace*{\fill}
\begin{minipage}[t]{6cm}
\centering
\includegraphics[width=6cm]{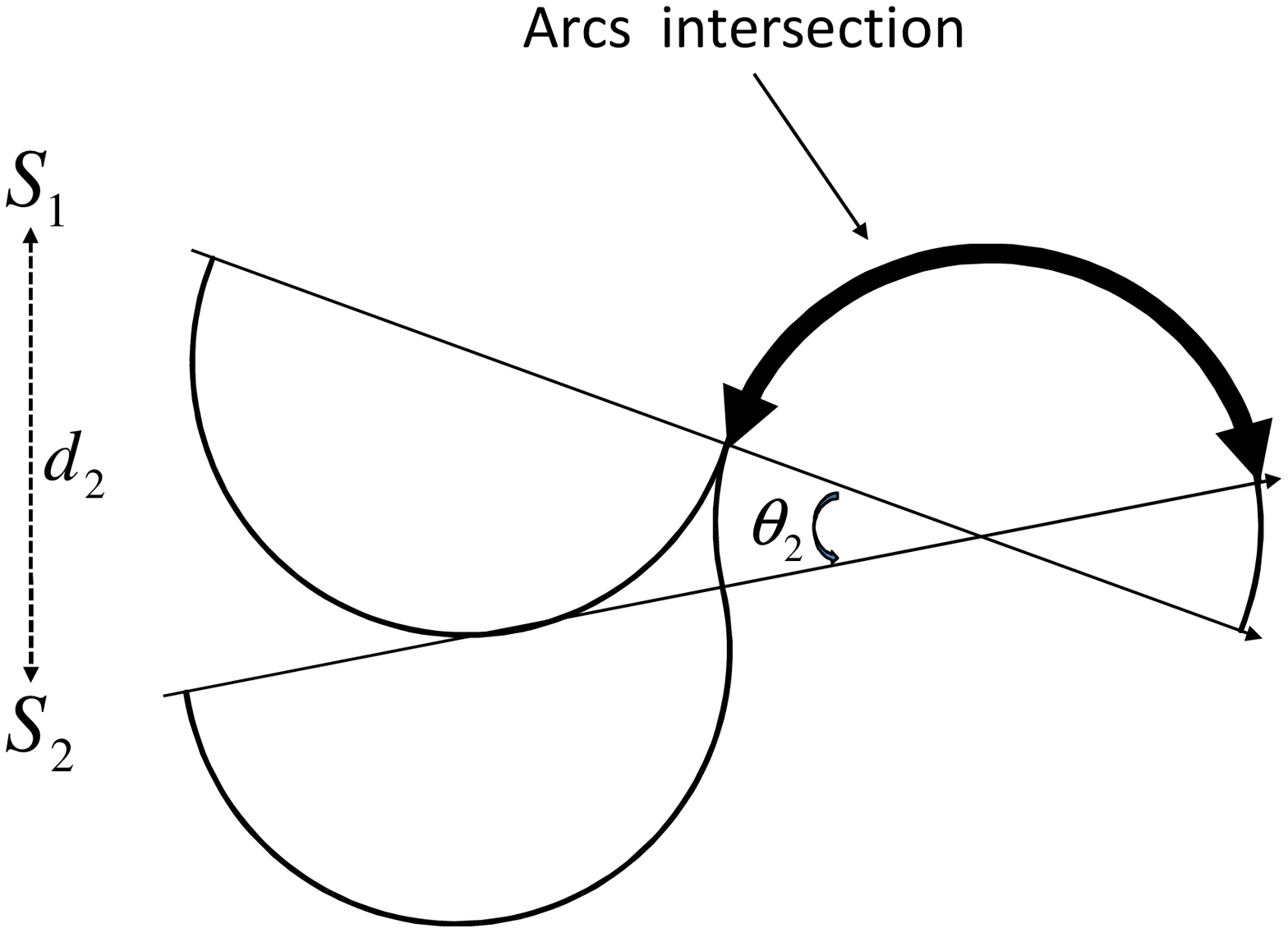}
\caption{\footnotesize Zoom on arcs intersection that produce interference pattern in the geodesics intersection region between slits screen and detector screen. Reduced geodesics arcs intersection for larger slits distance $d_2=d(S_1,S_2)$.}\label{Fig.8}
\end{minipage}
\end{figure}

\begin{figure}[!h]
\begin{minipage}[t]{6cm}
\centering
\includegraphics[width=6cm]{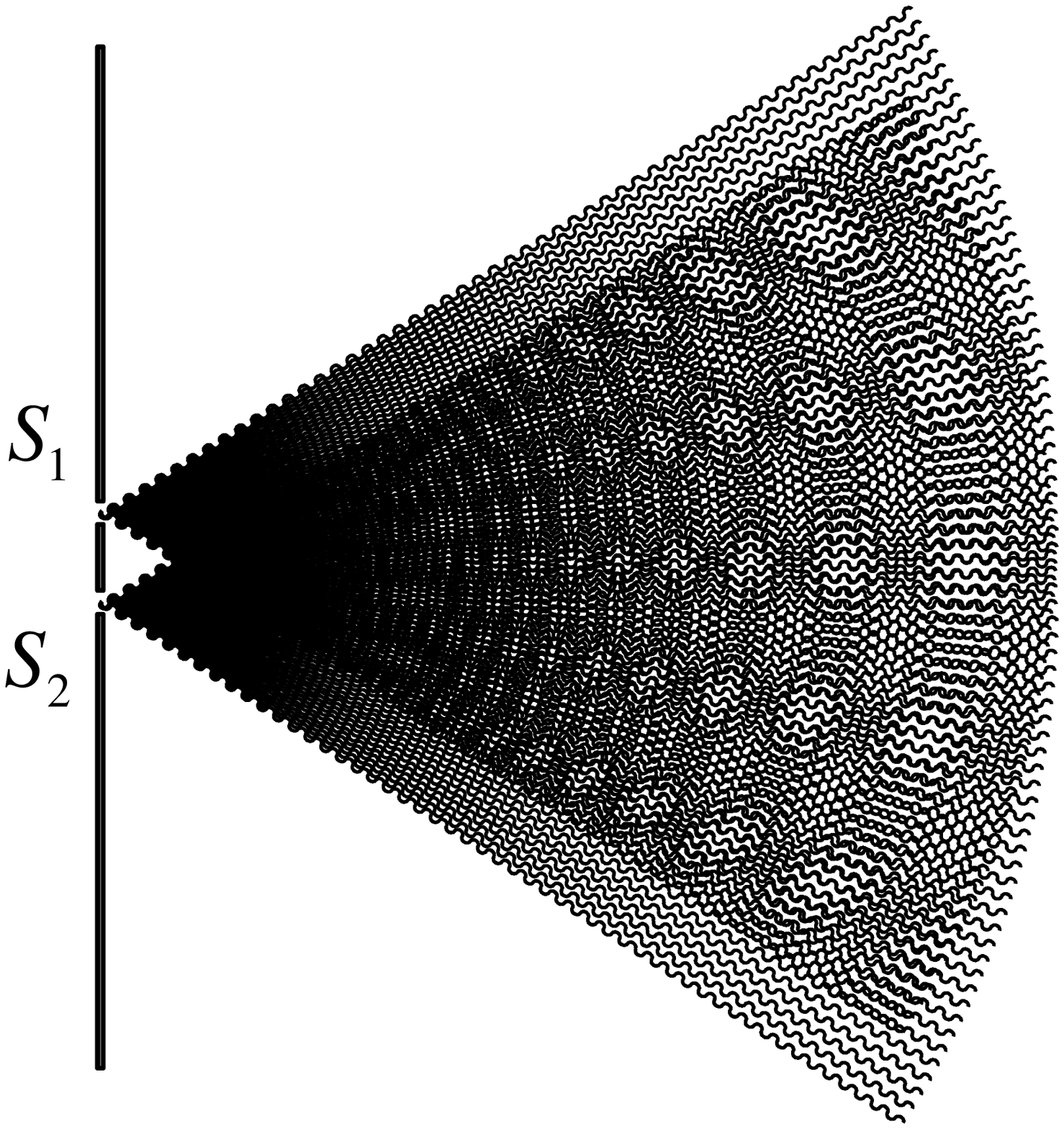}
\caption{\footnotesize Superposition of geodesics of radius $r_0=0.25mm$ from the slit $S_1$ with family of geodesics with radius $r_0=0.25mm$ from slit $S_2$ generate interference pattern in the geodesics intersection region. The number of clear pattern increases when the distance between the slits increases, $d(S_1,S_2)=2\ mm$.}\label{Fig.5}
\end{minipage}
\hspace*{\fill}
\begin{minipage}[t]{6cm}
\centering
\includegraphics[width=6cm]{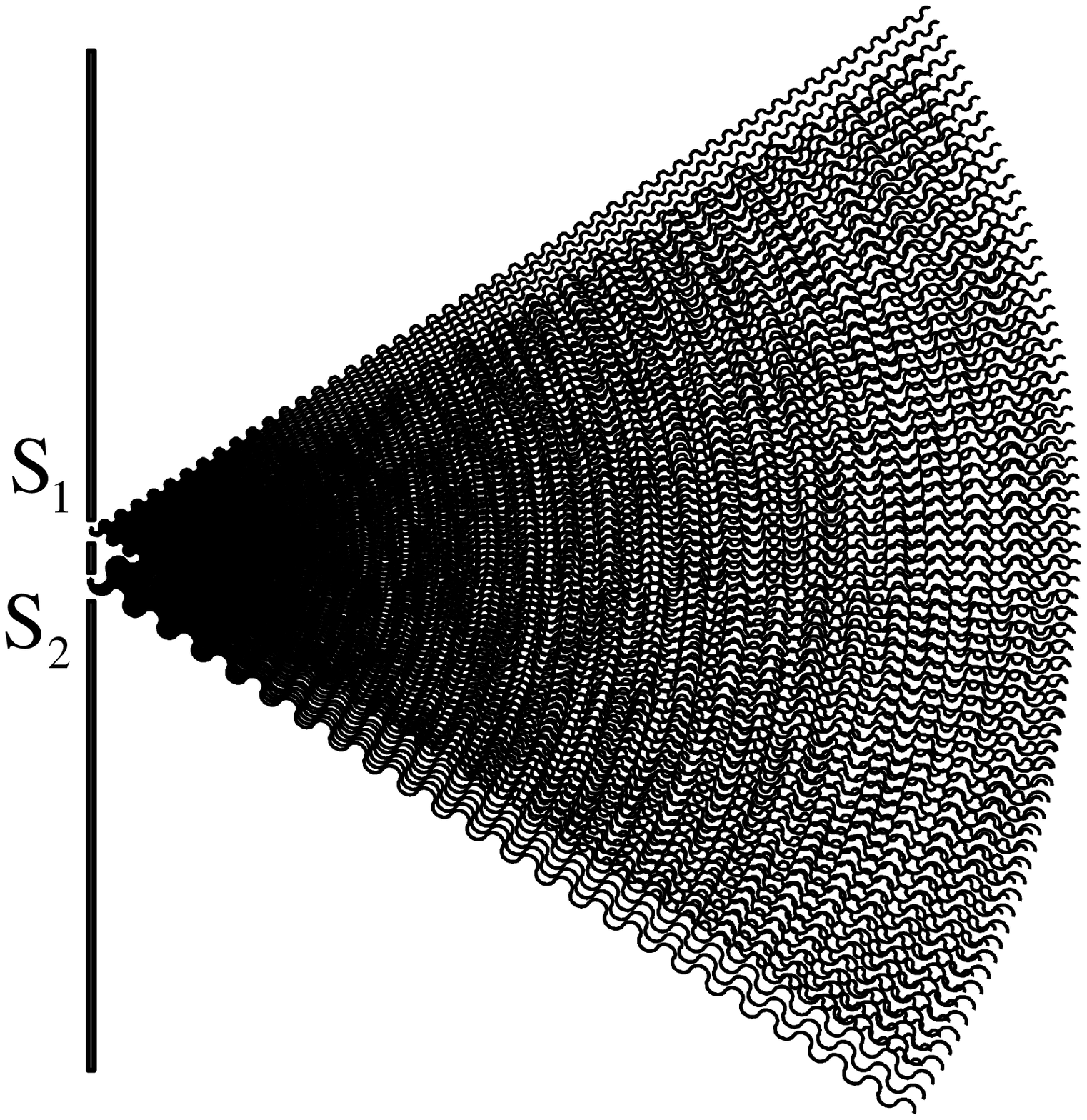}
\caption{\footnotesize The interference pattern disappears when we superimpose the family of geodesics of radius $r_0=0.25mm$ from the slit $S_1$ with a family of perturbed geodesics (paths with radius $r_1=0.625$ mm) from the slit $S_2$.}\label{Fig.6}
\end{minipage}
\end{figure}

The superimposition of two distant copies of the geodesic diffraction illustrated in (Fig.\ref{Fig.3}) simulates geodesics diffraction from two distant slits where their superimposition reproduces fringes and bands pattern observed in Young's double-slit experiment not only on a given detector screen but also within the whole geodesics intersection region between the slits and detector screen (Fig.\ref{Fig.4}).
The interference patterns in the geodesics intersection region appears because of the existence of sets of point intersections (arcs intersection Fig.\ref{Fig.7}). More precisely, if the local fluctuation centers coincide, there are arcs intersection that form the interference (clear spots in Fig.\ref{Fig.4}). Otherwise, there is only single point intersection.
The presence of the physical system is doubled within the intersection circle arcs that form the interference pattern.
Thus, at the clear spots region the probability of presence of the physical systems is doubled (in the 2D simulation) within circle of arcs intersection, meanwhile the probability of presence of the physical systems in the dark regions is doubled in single points of intersection, which creates the appearance of fringes and interference pattern within the intersection region as well as in the detector screen. Similarly in 3D, the clear spots mean that probability of presence of the physical systems is multiple within spheres surface intersection when the spheres centers coincide, meanwhile dark spots mean that the probability of presence of the physical systems is multiple within arcs of spheres intersection. In both cases the probability of presence of the physical systems in the intersection region between slits as well as in the detector screen is more or less important according to the existence of sets of geodesics intersection. Moreover, the closer the slits are, the larger the arcs intersection is (see Fig.\ref{Fig.7} and Fig.\ref{Fig.8}) and the number of fringes increases. Their size decreases when the distance between the slits $S_1$ and $S_2$ increases (see Fig.\ref{Fig.4} and Fig.\ref{Fig.5}).

The superimposition of diffracted geodesics (Fig.\ref{Fig.3}) from each slit allows to explain in geometrical optic why and how certain regions can be privileged for a physical system that follows these geodesics to build up fringe pattern, and in particular it explains the observed fringe pattern of electron, atoms and molecule when their motion follows paths of least time given by (Fig.\ref{Fig.00}), which suggests that the interference pattern is a pure manifestation of the space-time geodesics rather than a wave nature. The wave behavior of matter and radiation appears to be dictated by the space-time geodesics.

\subsection{Observation and final state perturbation}

Perturbation of the space-time geodesics provides an explanation of how observation affects the final state of the physical system. Indeed, the interference pattern totally disappears when we superimpose a diffracted family of geodesics with radius $r_0=0.25mm$ from the slit $S_1$ with a diffracted family of perturbed geodesics with bigger radius $r_1=0.625$ mm from the slit $S_2$. Their superimposition is illustrated in (Fig.\ref{Fig.6}), and it manifests absence of fringe in the intersection region. Perturbation in (Fig.\ref{Fig.6}) can be interpreted as follow:  after interaction with light (since experimental observations use light, and a photon or an electron cannot be detected without interaction with photons), the physical system will follow another path that is not a path with the lowest radius, it follows a geodesic with bigger radius that represents another path for the physical system state after interaction. The superimposition of two different families of geodesics with different radius of fluctuation that flare out from two distant slits will manifest absence of fringe in the intersection region as well as in the detector screen. A more precise simulation of superimposition of geodesics shows that total disorder occurs for a perturbation starting from $r_1 ={3\over 2} r_0$ as indicated in table \ref{T1}.
\begin{table}[!h]
\caption{Example of interference pattern from partial to total disorder}%%%Table caption goes here
\label{T1}
\begin{tabular}{llll}%%%The number of columns has to be defined here
\hline
Slit $S_1$ &Slit $S_2$&$d(S_1,S_2)$ &Interference \\
\hline
$r_0=1mm$ &$r_1=1.1mm$ &10mm  & appearance of disorder \\
$r_0=1mm$ &$r_1=1.2mm$ &10mm  & important disorder\\
$r_0=1mm$ &$r_1=1.3mm$ &10mm  & more important disorder \\
$r_0=1mm$ &$r_1=1.5mm$ &10mm  & total disorder\\\hline
\end{tabular}
\vspace*{-4pt}
\end{table}

\section{Mathematical formulation of infinity of paths of least time}
\subsection{Infinity of paths of least time in 2D}
The mathematical formulation of the geodesic illustrated in two dimensions (Fig.\ref{Fig.0}) and introduced in \cite{3} as paths of least time in an homogenous, isotropic and expanding space-time is given by the graph of the function $\varphi(x)$ defined for all $x\in \rR^+$ by:
\begin{equation}\label{Geod}
\varphi(x)=\sum_{n=0}^{N-1}g_{n}(x)
\end{equation}
where
\begin{equation}
g_{n}(x)=\left\{
             \begin{array}{ll}
               (-1)^n\ \sqrt{r^2 - \Big(x-(2n+1)r\Big)^2} & \hbox{for}\quad x\in [2nr,2(n+1)r]\\
               0 & \hbox{for}\quad x\not\in [2nr,2(n+1)r].
             \end{array}
           \right.
\end{equation}

For $n=0,\dots, N-1$, the graph of $g_{n}$ represents the geodesic between two antipodal points on the circle of center
$C_{n}=\Big((2n+1)r, 0\Big)$ and radius $r$. For $n=0,\dots, N-1$ the function $g_{n}$ is continuous on the closed interval $[2nr,2(n+1)r]$, differentiable on the open interval  $]2nr,2(n+1)r[$ and not differentiable at the points\quad $x_{n}=2nr$ and $x_{n+1}=2(n+1)r$.

\subsection{Infinity of paths of least time in 3D}
To provide a mathematical formulation of the infinity of paths of least time, the parametric equations of (\ref{Geod}) in 3D for $N$ aligned balls is given by:
\begin{equation}\label{Para}
\left\{
  \begin{array}{ll}
    x & =x \\
    y & = \varphi(x)\\
    z & =0
  \end{array}
\right.
\end{equation}
and a rotation of the geodesic (\ref{Para}) about the $x$-axis with angle $\theta\in[0,2\pi]$ is defined by the following parametric equations
\begin{equation}\label{Ro}
\left\{
  \begin{array}{ll}
    x' &= x \\
    y' & = \varphi(x)\cos\theta\\
    z' & =\varphi(x)\sin\theta
  \end{array}
\right.
\end{equation}
that defines the parametric equations of any geodesics in three dimensions illustrated in (Fig.\ref{Fig.00}). Any section of the revolution surface (\ref{Ro}) is a circle of equation defined by $y'^{2}+z'^{2}=\varphi^2(x)$ that gives for all $\theta\in[0,{2\pi}]$ the equation
\begin{equation}\label{sur}
y'=\pm\sqrt{\varphi^2(x) - z'^2},
\end{equation}
and by substitution of $z'$ in (\ref{sur}) we obtain
 \begin{equation}
 y'=\pm\sqrt{\varphi^2(x) - \varphi^2(x)\sin^2\theta}=\pm\varphi(x)\sqrt{1 - \sin^2\theta}=\pm\varphi(x)\cos\theta
 \end{equation}
 that defines two opposite fluctuating curves for all $\theta\in[0,{2\pi}]$, thus it is sufficient to use $y'=\varphi(x)\cos\theta$ for all $\theta\in[0,{2\pi}]$. The function (\ref{Geod}) is a plan function, and it remains a plan function for any rotation of angle $\theta\in[0,2\pi]$ with respect to the x-axis. For the following, we denote for all $x\in\rR^+$, for all $\theta\in[0,{2\pi}]$
\begin{equation}\label{Geod3d}
X(x,\theta):=\varphi(x)\cos\theta=\sum_{n=0}^{N-1}g_{n}(x,\theta)
\end{equation}
where $\varphi(x)$ is given by (\ref{Geod}), and for $n=0,\dots, N-1$
\begin{equation}\label{geod3d}
g_{n}(x,\theta):=\left\{
             \begin{array}{ll}
               (-1)^n\ \sqrt{r^2 - \Big(x-(2n+1)r\Big)^2}\cos\theta & \hbox{for}\quad x\in [2nr,2(n+1)r]\\
               0 & \hbox{for}\quad x\not\in [2nr,2(n+1)r].
             \end{array}
           \right.
\end{equation}

\subsection{Properties}

To parameterize the function (\ref{Geod3d}) with time it is sufficient to replace the parameter $x$ in the function (\ref{Geod3d}) by the function $x_t:=x(t)$ a continuous and differentiable function on $\rR^+$.
The geodesics (\ref{Geod3d}) are then given for all $\theta\in[0,{2\pi}]$, for all $t\in \rR^+$ by
\begin{equation}\label{Geod3dt}
X(x_t,\theta)=\sum_{n=0}^{N-1}g_{n}(x_t,\theta)
\end{equation}
where
\begin{equation}\label{geod3dt}
g_{n}(x_t,\theta)=\left\{
             \begin{array}{ll}
               (-1)^n\ \sqrt{r^2 - \Big(x_t-(2n+1)r\Big)^2}\cos\theta & \hbox{for}\quad x_t\in [2nr,2(n+1)r]\\
               0 & \hbox{for}\quad x_t\not\in [2nr,2(n+1)r]
             \end{array}
           \right.
\end{equation}

Consider a physical system of mass $m$ moving on the geodesics defined by (\ref{Geod3dt}), assuming that for all $t\in[t_a,t_b]\subset \rR^+$, $X(x(t_a),\theta)=a$ and $X(x(t_b),\theta)=b$ are two distant positions in the space-time. The motion of the physical system is completely unknown because of the existence of an infinity of paths of least time between the two distant positions.
Therefore for all $t\in[t_a,t_b]$, there exists $n\in\nN $ such that for all $x_t\in ]2nr,2(n+1)r[$, for all $\theta\in]0,{2\pi}[\backslash \{{\pi\over2},\pi, {3\pi\over2}\}$ the function $X(x(t),\theta)=g_n(x_t,\theta)$ is differentiable. For simplicity ,we denote $X_{x_t\theta}:=X(x(t),\theta)$ then for all $t\in]t_a,t_b[$  the instantaneous rate of change  of (\ref{Geod3dt}) is given by
\begin{equation}\label{vt}
V_t:={\partial X_{x_t\theta}\over\partial t}={\partial g_n(x_t,\theta)\over\partial t}=(-1)^{n+1} {(x_t-(2n+1)r)\over\sqrt{r^2 - \Big(x_t-(2n+1)r\Big)^2}}{\partial x_t\over\partial t}\cos\theta,
\end{equation}
meanwhile the changes of $X_{x_t\theta}$ in $\theta$ gives
\begin{equation}\label{vte}
V_\theta:={\partial X_{x_t\theta}\over\partial\theta}={\partial g_n(x_t,\theta)\over\partial\theta}=(-1)^{n+1}\sqrt{r^2 - \Big(x_t-(2n+1)r\Big)^2}\sin\theta.
\end{equation}

Then for all $\theta\in]0,{2\pi}[\backslash \{{\pi\over2},\pi, {3\pi\over2}\}$, and for $\beta:={\cos\theta\over\sin\theta}$, the equation (\ref{vte}) becomes
\begin{equation}\label{dx}
\beta {\partial X_{x_t\theta}\over\partial\theta}=-X_{x_t\theta},
\end{equation}
meanwhile the changes of $V_\theta$ in $\theta$ gives
\begin{equation}\label{dte}
{\partial V_t\over\partial\theta}=(-1)^{n} {(x_t-(2n+1)r) \over \sqrt{r^2 - \Big(x_t-(2n+1)r\Big)^2}}{\partial x_t\over\partial t}\sin\theta,
\end{equation}
and then
\begin{equation}\label{Majicc}
\beta{\partial V_t\over\partial\theta}=-V_t.
\end{equation}
\section{Action and wave function associated to the infinity of paths of least time}
\subsection{Action and wave function associated to each path of least time}

The use of the infinity of paths of least time for the quantum physical system motion between two different locations leads to associate to each path of least time a probability amplitude (following the essence of Feynman's path integral) that corresponds to the probability of finding the physical system in a certain location at a given angle $\theta\in]0,{2\pi}[\backslash \{{\pi\over2},\pi, {3\pi\over2}\}$. For this purpose we introduce the following: consider for all $t\in[t_a,t_b]$, for all $\theta\in]0,{2\pi}[\backslash \{{\pi\over2},\pi, {3\pi\over2}\}$ the action function defined by
\begin{eqnarray}\label{Actc}
   S:\rR\times [t_a,t_b]& \rightarrow &\rR\nonumber \\
        \Big(X_{x_t\theta}, t\Big) &\mapsto & S(X_{x_t\theta},t)
\end{eqnarray}
related to the momentum, total energy and angle variation by
\begin{eqnarray}
  {\partial{S}\over\partial {X_{x_t\theta}}} &=&m{V}_{t}= {P}\label{Act1c} \\
  -{\partial{S}\over\partial t} &=&E\label{Act2c} \\
  {\partial{S}\over\partial \theta}&=& {-\hbar\over 2\beta}.\label{Act3c}
\end{eqnarray}

Consider for all $\theta\in]0,{2\pi}[\backslash \{{\pi\over2},\pi, {3\pi\over2}\}$, for all $x_t\in [2nr,2(n+1)r]$ the  wave function associated to each path defined in (\ref{Geod3dt}) by
\begin{eqnarray}\label{Wc}
   \Psi:\rR\times [t_a,t_b]& \rightarrow &\cC\nonumber \\
        (X_{x_t\theta}, t) &\mapsto & \Psi(X_{x_t\theta}, t)=e^{{i\over \hbar}{S}(X_{x_t\theta}, t)},
\end{eqnarray}
where $i$ is the imaginary unit, $\hbar={h\over2\pi}$ and $h$ is the Plank's constant. The function $\Psi(X_{x_t\theta}, t)$ represents the wave function for a single particle where $X_{x_t\theta}$ is its position in space at the time $t\in[t_a,t_b]$ and the angle $\theta\in]0,{2\pi}[\backslash \{{\pi\over2},\pi, {3\pi\over2}\}$ (this wave function was used by Feynman in his postulate II of the path integral). Then similarly to Feynman's choice, the  wave function (\ref{Wc}) represents a probability amplitude that corresponds to the probability of finding the physical system in a certain location at the time $t$, and the angle $\theta\in]0,{2\pi}[\backslash \{{\pi\over2},\pi, {3\pi\over2}\}$. The wave function $\Psi(X_{x_t\theta}, t)$ may take the approximate form
$\Psi(X_{x_t\theta}, t)=Ae^{{i\over \hbar}{\cal S}(X_{x_t\theta}, t)}$
where $A$ is the dimensional constant required for any physical interpretation. In the following the amplitude $A$ of the used wave function will be considered as equal to 1 for more simplicity and it can be reintroduced if needed in any linear equation. The wave function (\ref{Wc}) gives, using equations (\ref{Act1c}) and (\ref{Act2c}), the correspondence principles of quantum mechanics for momentum and total energy for all $x_t\in ]2nr,2(n+1)r[$, for all $\theta\in]0,{2\pi}[\backslash \{{\pi\over2},\pi, {3\pi\over2}\}$,
\begin{equation}\label{Cpc}
{P}=-i\hbar {\partial\Psi(X_{x_t\theta},t)\over\partial X_{x_t\theta}}{1\over \Psi(X_{x_t\theta},t)},\qquad\hbox{and}\qquad E=i\hbar {\partial\Psi(X_{x_t\theta},t)\over\partial t}{1\over \Psi(X_{x_t\theta},t)}.
\end{equation}

\subsection{Compatibility with the indeterminacy principle}

The wave function (\ref{Wc}) is interpreted as a probability amplitude that represents the wave function for a single particle, in which $X_{x_t\theta}$ is its position in space at the time $t\in[t_a,t_b]$, and angle $\theta\in]0,{2\pi}[\backslash \{{\pi\over2},\pi, {3\pi\over2}\}$ and where the square modulus of the wave function
\begin{equation}\label{ModL}
\vert\Psi(X_{x_t\theta},t)\vert^2=\Psi(X_{x_t\theta},t)\Psi^*(X_{x_t\theta},t)=\rho(X_{x_t\theta},t),
 \end{equation}
for $\Psi^*$ the conjugate of $\Psi$, represents the probability density that the physical system at the time $t\in[t_a,t_b]$ and angle $\theta\in]0,{2\pi}[\backslash \{{\pi\over2},\pi, {3\pi\over2}\}$ is located at $X_{x_t\theta}$.
If one measures the position of the physical system at the time $t\in[t_a,t_b]$ and the angle $\theta\in]0,{2\pi}[\backslash \{{\pi\over2},\pi, {3\pi\over2}\}$, then the location of the physical system is expressed by the probability distribution as follow: the probability that the physical system's position $X_{x_t\theta}$ is between $X_{x_t\theta_0}=c$ and $X_{x_t\theta_1}=d$, is given by the following integral of density
\begin{equation}\label{Prob01L}
P_{c\leq X_{x_t\theta}\leq d}(t)=\int^d_c \vert\Psi(X_{x_t\theta},t)\vert^2 dX_{x_t\theta}=\int^{\theta_1}_{\theta_0} \vert\Psi(X_{x_t\theta},t)\vert^2 {\partial X_{x_t\theta}\over\partial \theta}d\theta,
\end{equation}
where $t$ and $\theta$ are the time and angle at which the physical system is measured. The normalization condition is then
\begin{equation}
\int^{+\infty}_{-\infty} \vert\Psi(X_{x_t\theta},t)\vert^2 dX_{x_t\theta}=\int^{2\pi}_{0} \vert\Psi(X_{x_t\theta},t)\vert^2 {\partial X_{x_t\theta}\over\partial \theta}d\theta=1.
\end{equation}

\subsection{Uncertainty relation, angles and action function}

The angle rate of change of the action function (\ref{Actc}) for all $\theta\in]0,{2\pi}[\backslash \{{\pi\over2},\pi, {3\pi\over2}\}$ given by (\ref{Act3c}) conveys an uncertainty relation. Indeed, based on equation (\ref{Act3c}) we have the following:
\begin{lem}\label{Lem1c}
 Let $S$ be the action function defined by (\ref{Actc}), and $\Psi$ the wave function defined by (\ref{Wc}). For all $(t,\theta)\in [t_a,t_b]\times ]0,{2\pi}[\backslash \{{\pi\over2},\pi, {3\pi\over2}\}$ we have
\begin{equation}\label{Heizc}
{\partial{ S}\over\partial \theta} = {-\hbar\over 2\beta}\quad\Longleftrightarrow\quad {P}\ X_{x_t\theta}={\hbar\over 2}
\end{equation}
where $X_{x_t\theta}$ is the position function, and ${P}=m{V}_{t}$ is the momentum.
\end{lem}
{\it Proof.} For all $(t,\theta)\in [t_a,t_b]\times ]0,{2\pi}[\backslash \{{\pi\over2},\pi, {3\pi\over2}\}$, $\exists n\in\nN$ such that $x_t\in ]2nr,2(n+1)r[$, and  $X_{x_t\theta}=g_n(x_t,\theta)$.
Thus, using definition (\ref{Wc}) of $\Psi$, ${S}(X_{x_t\theta},t)=-i\hbar\ln\Psi(X_{x_t\theta},t)$,
and the chain rule with (\ref{Act3c}) gives
\begin{equation}\label{Cha1c}
{\partial S\over\partial \theta}={\partial S\over\partial X_{x_t\theta}} {\partial X_{x_t\theta}\over\partial\theta}={-\hbar\over 2\beta},
\end{equation}
then by substitution of (\ref{Act1c}) and (\ref{vte}) in (\ref{Cha1c}), we obtain $mV_t V_\theta={-\hbar\over 2\beta}$, and the use of equation (\ref{dx}) gives the result ${P}\ X_{x_t\theta} = {\hbar\over2}$.

\section{Derivation of Schr\"{o}dinger Equation}
\subsection{Infinity of paths of least time in complex plan}

For the derivation of Schr\"{o}dinger equation, the infinity of geodesics needs to be defined in the complex set $\cC$. Thus, consider for all $t\in[t_a,t_b]$, for all $\theta\in]0,{2\pi}[\backslash \{{\pi\over2},\pi, {3\pi\over2}\}$ the complex geodesics $u_{x_t\theta}:=kX_{x_t\theta}$ where $k\in\cC$ is a complex number that verifies $k^2=i$ and $X_{x_t\theta}$ are the space-time geodesics (\ref{Geod3dt}) introduced above. Then using (\ref{vt}) and (\ref{vte}) we have for all $(t,\theta)\in [t_a,t_b]\times ]0,{2\pi}[\backslash \{{\pi\over2},\pi, {3\pi\over2}\}$,
\begin{equation}\label{Cvt}
\mathbb{V}_t:={\partial  u_{x_t\theta}\over\partial t}=kV_t, \qquad\hbox{and}\qquad \mathbb{V}_\theta:={\partial u_{x_t\theta}\over\partial\theta}=kV_\theta,
\end{equation}
meanwhile using (\ref{dx}) for $\beta={\cos\theta\over\sin\theta}$, $\theta\in]0,{2\pi}[\backslash \{{\pi\over2},\pi, {3\pi\over2}\}$, we have
\begin{equation}\label{dx2}
\beta {\partial u_{x_t\theta}\over\partial\theta}=-u_{x_t\theta},
\end{equation}
and the use of (\ref{Majicc}) gives
\begin{equation}\label{MagiC}
\beta{\partial \mathbb{V}_t\over\partial\theta}=-\mathbb{V}_t.
\end{equation}

\subsection{Action function, wave function and uncertainty in the complex plan}

Consider for all $(t,\theta)\in [t_a,t_b]\times ]0,{2\pi}[\backslash \{{\pi\over2},\pi, {3\pi\over2}\}$ the action function defined  by
\begin{eqnarray}\label{cAct}
   \mathbb{S}:\cC\times [t_a,t_b]& \rightarrow &\cC\nonumber \\
        (u_{x_t\theta},t) &\mapsto & \mathbb{S}(u_{x_t\theta},t)=S(kX_{x_t\theta},t),
\end{eqnarray}
related to the momentum, total energy and angle variation by
\begin{eqnarray}
  {\partial \mathbb{S}\over\partial u_{x_t\theta}} &=&m\ \mathbb{V}_{t}=  \mathbb{P}\label{cAct1} \\
  -{\partial \mathbb{S}\over\partial t} &=& \mathbb{E} \label{cAct2} \\
  {\partial \mathbb{S}\over\partial \theta} &=&-{i\hbar\over 2\beta}.\label{cAct3}
\end{eqnarray}

Consider for all $(t,\theta)\in [t_a,t_b]\times ]0,{2\pi}[\backslash \{{\pi\over2},\pi, {3\pi\over2}\}$, the wave function defined by
\begin{eqnarray}\label{cW}
   \Psi:\cC\times [t_a,t_b]& \rightarrow &\cC\nonumber \\
        (u_{x_t\theta},t) &\mapsto & \Psi(u_{x_t\theta},t)=e^{{i\over \hbar} \mathbb{S}(u_{x_t\theta},t)}.
\end{eqnarray}

Since $u_{x_t\theta}=kX_{x_t\theta}$, then the function $\Psi(u_{x_t\theta},t)=\Psi(kX_{x_t\theta},t)$ represents the wave function associated to the physical system of mass $m$ where $X_{x_t\theta}$ is its position at the time $t$ and the angle $\theta$. The squared module $\vert{\Psi}(u_{x_t\theta},t)\vert^2$ represents the probability density for the particle at time $t$ and the angle $\theta\in]0,{2\pi}[\backslash \{{\pi\over2},\pi, {3\pi\over2}\}$ to be in the position $X_{x_t\theta}$. The probability that the physical system's position $X_{x_t\theta}$ is between $X_{x_t\theta_0}=c$ and $X_{x_t\theta_1}=d$ is given by
\begin{equation}\label{cProb01L}
P_{c\leq X_{x_t\theta}\leq d}(t)=\int^d_c \vert\Psi(u_{x_t\theta},t)\vert^2 dX_{x_t\theta}=\int^{\theta_1}_{\theta_0} \vert\Psi(u_{x_t\theta},t)\vert^2 {\partial X_{x_t\theta}\over\partial \theta}d\theta,
\end{equation}
where $t$ and $\theta$ are the time and the angle at which the physical system is measured. The normalization condition is given by
\begin{equation}
\int^{+\infty}_{-\infty} \vert\Psi(u_{x_t\theta},t)\vert^2 dX_{x_t\theta}=\int^{2\pi}_{0} \vert\Psi(u_{x_t\theta},t)\vert^2 {\partial X_{x_t\theta}\over\partial \theta}d\theta=1.
\end{equation}

The wave function (\ref{cW}) gives, using equation (\ref{cAct1}) and (\ref{cAct2}), the correspondence principle of quantum mechanics for momentum and total energy
\begin{equation}\label{cCp}
\mathbb{P}=-i\hbar {\partial\Psi(u_{x_t\theta},t)\over\partial u_{x_t\theta}}{1\over \Psi(u_{x_t\theta},t)},\qquad\hbox{and}\qquad \mathbb{E}=i\hbar {\partial\Psi(u_{x_t\theta},t)\over\partial t}{1\over \Psi(u_{x_t\theta},t)}.
\end{equation}

The action functions (\ref{Actc}) and (\ref{cAct}) for all $(t,\theta)\in [t_a,t_b]\times ]0,{2\pi}[\backslash \{{\pi\over2},\pi, {3\pi\over2}\}$, verify the following equalities  ${\partial \mathbb{S}\over\partial u_{x_t\theta}} = k\ {\partial {S}\over\partial X_{x_t\theta}}$,
${\partial \mathbb{S}\over\partial t}= i\ {\partial {S}\over\partial t}$, and  ${\partial \mathbb{S}\over\partial \theta} = i\ {\partial {S}\over\partial \theta}$. Indeed from equation (\ref{cAct1}) and (\ref{Cvt}), we have
${\partial \mathbb{S}\over\partial u_{x_t\theta}}=\mathbb{P}=m\mathbb{V}_{t}=mk{V}_{t}=k{\partial {S}\over\partial X_{x_t\theta}}$.
The chain rule and the use of $u_{x_t\theta}=kX_{x_t\theta}$ give ${\partial \mathbb{S}\over\partial t}={\partial \mathbb{S}\over\partial u_{x_t\theta}} {\partial u_{x_t\theta}\over\partial t}=k{\partial {S}\over\partial X_{x_t\theta}} k {\partial X_{x_t\theta}\over\partial t}=i{\partial{S}\over\partial t}$.
The equality ${\partial \mathbb{S}\over\partial \theta} = i\ {\partial {S}\over\partial \theta}$ is evident using equalities (\ref{cAct3}) and (\ref{Act3c}).
\begin{lem}\label{Lem2}
Let $\mathbb{S}$ be the action function defined by (\ref{cAct}), and $\Psi$ the wave function defined by (\ref{cW}), then for all $(t,\theta)\in [t_a,t_b]\times ]0,{2\pi}[\backslash \{{\pi\over2},\pi, {3\pi\over2}\}$, we have
\begin{equation}\label{CHeiz}
{\partial \mathbb{S}\over\partial\theta} =-{i\hbar\over2\beta}\quad\Longleftrightarrow\quad  \mathbb{P}\ u_{x_t\theta}= {i\hbar\over 2},
\end{equation}
where $u_{x_t\theta}$ is the virtual position function (the position function multiplied by a complex number $k$ such that $k^2=i$), and where $\mathbb{P}=m{\mathbb V}_{t}=k\ (mV_t)$ the virtual momentum.
\end{lem}
{\it Proof:} Using equality (\ref{cAct3}) together with ${\partial \mathbb{S}\over\partial \theta} = i\ {\partial {S}\over\partial \theta}$ for all $(t,\theta)\in [t_a,t_b]\times ]0,{2\pi}[\backslash \{{\pi\over2},\pi, {3\pi\over2}\}$, gives
$i{\partial {S}\over\partial\theta} =-{i\hbar\over2\beta}$, then from Lemma \ref{Lem1c} we have ${P}\ X_{x_t\theta}={\hbar\over 2}$, and the multiplication by $k^2$ allows to conclude.

\subsection{Infinity of paths of least time and derivation of Schr\"odinger's equation}

Using the wave function (\ref{cW}) we have the following:
\begin{thm}\label{cTSch}
Let $\mathbb{S}$ be the action function defined by (\ref{cAct}) that verifies equations (\ref{cAct1}), (\ref{cAct2}), and (\ref{cAct3}), and let $\Psi$ be the wave function defined by (\ref{cW}), then for all $(t,\theta)\in [t_a,t_b]\times ]0,{2\pi}[\backslash \{{\pi\over2},\pi, {3\pi\over2}\}$, the wave function $\Psi$ satisfies the time evolution equation:
\begin{equation}\label{CSchrod}
i{\hbar}{\partial\Psi(u_{x_t\theta},t)\over\partial t}+{\hbar^2\over 2m}{\partial^2\Psi(u_{x_t\theta},t)\over\partial u_{x_t\theta}^2}=\Phi(u_{x_t\theta},t)\Psi(u_{x_t\theta},t),
\end{equation}
where the potential function is given by $\Phi(u_{x_t\theta},t)=-{\mathbb{P}^2\over 2m}$.
\end{thm}
{\it Proof.} We denote $\Psi:=\Psi(u_{x_t\theta},t)$ for simplicity, for all $t\in[t_a,t_b]$ there exists $n\in\nN$ such that $x_t\in[2nr,2(n+1)r]$, then for all $x_t\in ]2nr,2(n+1)r[$ and for all $\theta\in]0,{2\pi}[\backslash \{{\pi\over2},\pi, {3\pi\over2}\}$ the equation  (\ref{cCp}) for momentum gives
\begin{equation}\label{cCV}
\mathbb{V}_{t}=-{i\hbar\over m}\ {\partial\Psi\over\partial u_{x_t\theta}}{1\over \Psi},
\end{equation}
and since $\mathbb{V}_{t}$  verify equation (\ref{MagiC}), then the substitution of equations (\ref{cCV}) in equation (\ref{MagiC}) gives
\begin{equation}\label{cM1}
{i\hbar\over m} {\partial\Psi\over\partial u_{x_t\theta}}{1\over \Psi}=-{i\hbar\over m}\beta\Big(\frac{\partial}{\partial\theta}\Big({\partial\Psi\over\partial u_{x_t\theta}}{1\over \Psi}\Big)\Big).
\end{equation}

The chain rule gives
\begin{equation}\label{cCa}
{\partial\Psi\over\partial u_{x_t\theta}}={\partial\Psi\over\partial t}{\partial t\over\partial u_{x_t\theta}}={\partial\Psi\over\partial t}{1\over  \mathbb{V}_{t}}
\end{equation}
and
\begin{equation}\label{cCb}
\frac{\partial}{\partial\theta}\Big({\partial\Psi\over\partial u_{x_t\theta}}{1\over \Psi}\Big)=\frac{\partial}{\partial u_{x_t\theta}}\Big({\partial\Psi\over\partial u_{x_t\theta}}{1\over \Psi}\Big)\frac{\partial u_{x_t\theta}}{\partial \theta}=\Big({\partial^2\Psi\over\partial u_{x_t\theta}^2}{1\over \Psi}-\Big({\partial\Psi\over\partial u_{x_t\theta}}\Big)^2{1\over \Psi^2}\Big)\frac{\partial u_{x_t\theta}}{\partial \theta}.
\end{equation}

Using equations (\ref{cCa}) and (\ref{cCb}) in equation (\ref{cM1}) gives
\begin{equation}\label{cM2}
{i\hbar\over m} {1\over \Psi}{\partial\Psi\over\partial t}{1\over  \mathbb{V}_{t}}=-{i\hbar\over m}\beta\Big({\partial^2\Psi\over\partial u_{x_t\theta}^2}{1\over \Psi}-\Big({\partial\Psi\over\partial u_{x_t\theta}}\Big)^2{1\over \Psi^2}\Big)\frac{\partial u_{x_t\theta}}{\partial \theta}.
\end{equation}

Using equation (\ref{dx2}) in equation (\ref{cM2}), we obtain
\begin{equation}\label{cM3}
{i\hbar\over m} {1\over \Psi}{\partial\Psi\over\partial t}{1\over  \mathbb{V}_{t}}=-{i\hbar\over m}\Big({\partial^2\Psi\over\partial u_{x_t\theta}^2}{1\over \Psi}-\Big({\partial\Psi\over\partial u_{x_t\theta}}\Big)^2{1\over \Psi^2}\Big) (-u_{x_t\theta}).
\end{equation}

The multiplication of both sides of the equation (\ref{cM3}) by $m\Psi \mathbb{V}_{t}$ together with the consideration of equation (\ref{cCp}) for momentum gives
\begin{equation}\label{cM4}
i\hbar{\partial\Psi\over\partial t}=i\hbar\Big({\partial^2\Psi\over\partial u_{x_t\theta}^2}-{\mathbb{P}^2\over(i\hbar)^2}\Psi\Big) \mathbb{V}_{t}\ (u_{x_t\theta}).
\end{equation}

Lemma \ref{Lem2} gives
\begin{equation}\label{cM5}
i\hbar{\partial\Psi\over\partial t}=i\hbar\Big({\partial^2\Psi\over\partial u_{x_t\theta}^2}-{\mathbb{P}^2\over(i\hbar)^2}\Psi\Big) ({i\hbar\over2m}),
\end{equation}
that is to say
\begin{equation}\label{cM7}
i{\hbar}{\partial\Psi\over\partial t}+{\hbar^2\over 2m}\Big({\partial^2\Psi\over\partial u_{\delta\varepsilon}^2}\Big)=-{\mathbb{P}^2\over 2m} \Psi,
\end{equation}
which concludes the proof.

\section{Zero point energy, Van Der Waals torque and Casimir effect}

To understand causality between energy variation and geometry variation of paths of least time, consider a physical system that has a weak interaction with matter, with non zero mass $m$, zero charge and that generates interference pattern in the Youngs double-slit experiment following the infinity of paths of least time (\ref{Geod3d}) for all $\theta\in[0,{2\pi}]$ (as example of physical system: neutrino). Let $\mathbb{S}$ be the action function defined by (\ref{cAct}) that verifies equations (\ref{cAct1}), (\ref{cAct2}), and (\ref{cAct3}), and let $\Psi$ be the wave function defined by (\ref{cW}), that satisfies for all $(t,\theta)\in [t_a,t_b]\times ]0,{2\pi}[\backslash \{{\pi\over2},\pi, {3\pi\over2}\}$, the time evolution equation (\ref{CSchrod}).

\subsection{Zero point energy and energy fluctuation}

The Schr\"{o}dinger equation (\ref{CSchrod}) provides interesting information about the zero point energy. Indeed
\begin{equation}\label{CSchrodP}
i{\hbar}{\partial\Psi(u_{x_t\theta},t)\over\partial t}+{\hbar^2\over 2m}{\partial^2\Psi(u_{x_t\theta},t)\over\partial u_{x_t\theta}^2}=-{\mathbb{P}^2\over 2m}\Psi(u_{x_t\theta},t),
\end{equation}
using the correspondence principle of energy (\ref{cCp}) it becomes
\begin{equation}\label{CSchrodP1}
E \Psi(u_{x_t\theta},t)=-{\hbar^2\over 2m}{\partial^2\Psi(u_{x_t\theta},t)\over\partial u_{x_t\theta}^2}-{\mathbb{P}^2\over 2m}\Psi(u_{x_t\theta},t),
\end{equation}
where the partial derivative of the product gives
\begin{equation}\label{CSchrodP2}
{\partial^2\Psi(u_{x_t\theta},t)\over\partial u_{x_t\theta}^2}={i\over\hbar}{\partial\over\partial u_{x_t\theta}}\Big({\partial\mathbb{S}\over\partial u_{x_t\theta}}\ \Psi(u_{x_t\theta},t)\Big)=\Big({i\over\hbar}{\partial^2\mathbb{S}\over\partial u_{x_t\theta}^2}+({i\over\hbar})^2\Big({\partial\mathbb{S}\over\partial u_{x_t\theta}}\Big)^2\Big)\Psi(u_{x_t\theta},t).
\end{equation}

The substitution of (\ref{CSchrodP2}) in (\ref{CSchrodP1}) gives
\begin{equation}\label{CSchrodP3}
E \Psi(u_{x_t\theta},t)=\Big({-i\hbar\over2m}{\partial^2\mathbb{S}\over\partial u_{x_t\theta}^2}+{1\over2m}\Big({\partial\mathbb{S}\over\partial u_{x_t\theta}}\Big)^2-{\mathbb{P}^2\over 2m}\Big)\Psi(u_{x_t\theta},t),
\end{equation}
and the substitution of equation (\ref{cAct1}) in (\ref{CSchrodP3}) gives
\begin{equation}\label{CSchrodP4}
E \Psi(u_{x_t\theta},t)=\Big({-i\hbar\over2m}{\partial^2\mathbb{S}\over\partial u_{x_t\theta}^2}+{\mathbb{P}^2\over 2m}-{\mathbb{P}^2\over 2m}\Big)\Psi(u_{x_t\theta},t)=\Big({-i\hbar\over2m}{\partial^2\mathbb{S}\over\partial u_{x_t\theta}^2}\Big)\Psi(u_{x_t\theta},t),
\end{equation}
that is to say
\begin{equation}\label{CSchrodP5}
E ={-i\hbar\over2m}{\partial^2\mathbb{S}\over\partial u_{x_t\theta}^2}.
\end{equation}

The chain rule as well as the use of equation (\ref{cAct1}), (\ref{dx2}), (\ref{MagiC}), and (\ref{CHeiz}) lead to find
\begin{equation}\label{CSchrodP6}
{\partial^2\mathbb{S}\over\partial u_{x_t\theta}^2}=m{\partial\mathbb{V}_t\over\partial u_{x_t\theta}}=m{\partial\mathbb{V}_t\over\partial\theta}{\partial\theta\over\partial u_{x_t\theta}}={\mathbb{P}\over u_{x_t\theta}}={i\hbar\over 2u_{x_t\theta}^2}
\end{equation}
and the substitution of equation (\ref{CSchrodP6}) in equation (\ref{CSchrodP5}) gives
\begin{equation}\label{CSchrodP9}
E ={\hbar^2\over4m u_{x_t\theta}^2}={\hbar^2\over4m k^2 X_{x_t\theta}^2}=-i{\hbar^2\over4m X_{x_t\theta}^2}=i\vert E\vert,
\end{equation}
where the energy magnitude of the physical system is given by
\begin{equation}\label{CSch}
\vert E\vert ={\hbar^2\over4m X_{x_t\theta}^2}.
\end{equation}

\subsubsection{Variation of the energy magnitude of the physical system}

\begin{figure}[!h]
\begin{minipage}[t]{6cm}
\centering
\includegraphics[width=6cm]{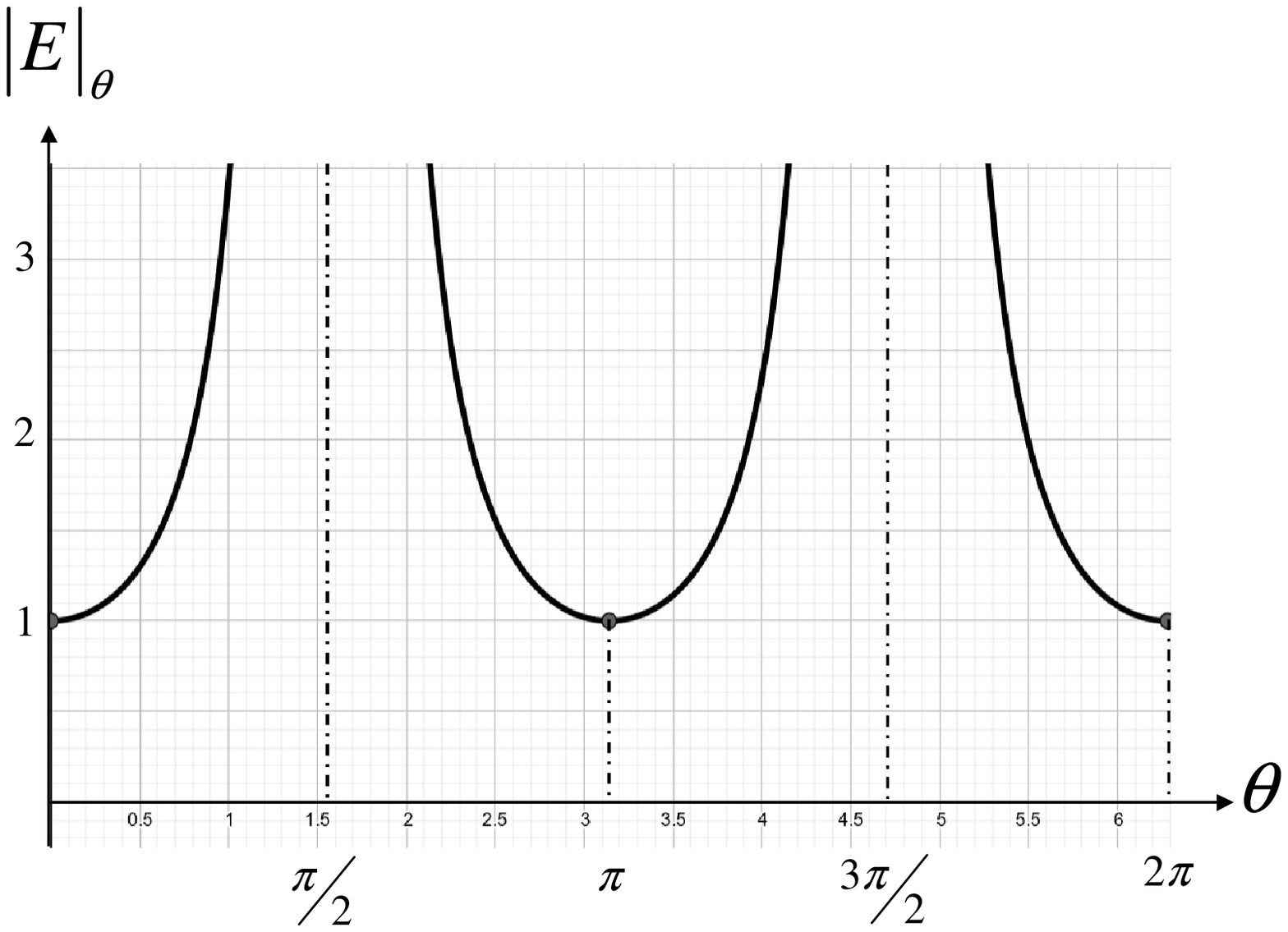}
\caption{\footnotesize Variation of the function $\vert E\vert_\theta={k\over \cos^2\theta}$, k=1, for all $\theta\in ]0,{2\pi}[\backslash \{{\pi\over2}, \pi, {3\pi\over2}\}$. For a small constant $k={\hbar^2\over 4mr^2}$ the graph of $\vert E\vert_\theta$ for all $\theta\in ]0,{2\pi}[\backslash \{{\pi\over2}, \pi, {3\pi\over2}\}$ remain convex but the minimums will be more small.}\label{Fig.13}
\end{minipage}
\hspace*{\fill}
\begin{minipage}[t]{6cm}
\centering
\includegraphics[width=6cm]{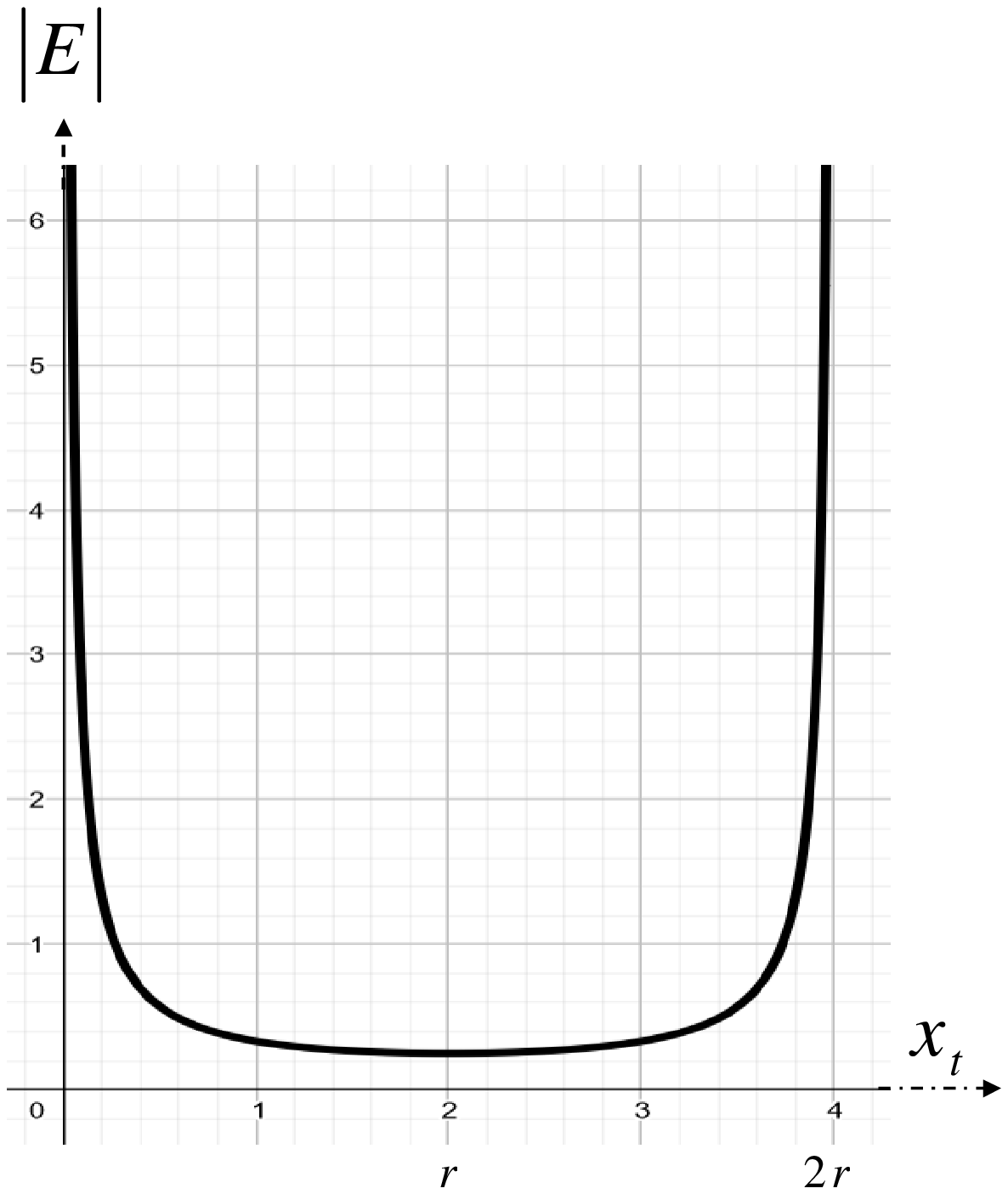}
\caption{\footnotesize Variation of the energy $\vert E\vert={k\over 2^2-(x_t-2)^2}$ with k=1 function of $x_t\in]0,4[$. The graph of $\vert E\vert$ defined in (\ref{CSch}) for $\theta=0$, $k={\hbar^2\over 4m}$ is convex function and its graph is similar to this graph.}\label{Fig.14}
\end{minipage}
\end{figure}

The energy magnitude (\ref{CSch}) is defined for all $(x_t,\theta)\in ]2nr,(2n+2)r[\times]0,{2\pi}[\backslash \{{\pi\over2},\pi, {3\pi\over2}\}$, varies as a convex function for $x_t\in ]2nr,(2n+2)r[$ (see Fig.\ref{Fig.14}) and varies as a convex function for $\theta\in]0,{2\pi}[\backslash \{{\pi\over2},\pi, {3\pi\over2}\}$ (see Fig.\ref{Fig.13}).
Indeed,

\textbf{i)} The variation of energy magnitude (\ref{CSch}) for $x_t\in ]2nr,(2n+2)r[$ is related to the paths of least time fluctuation (it increases near the antipodal points $(2nr,0,0)$ and $(2n+2)r,0,0)$ and is minimal at the center of fluctuation $(2n+1)r,0,0)$ for all $n\in\nN$). This energy variation function of  $x_t\in ]2nr,(2n+2)r[$ characterizes the motion of the physical system on the fluctuating paths of least time and since energy cannot be created or destroyed, this variation is due to the continuous transfer of part of energy of the physical system into changes of motion direction on each path of least time in the path plan.

\textbf{ii)} The variation of energy magnitude (\ref{CSch}) for $\theta\in ]2nr,(2n+2)r[\times]0,{2\pi}[\backslash \{{\pi\over2},\pi, {3\pi\over2}\}$ is related to the change of paths of least time function of $\theta$ (polarization). It is a transversal variation of the energy magnitude (\ref{CSch}) in the plan perpendicular to the paths of least time principal axis.

\begin{figure}[!h]
\begin{minipage}[t]{6cm}
\centering
\includegraphics[width=6cm]{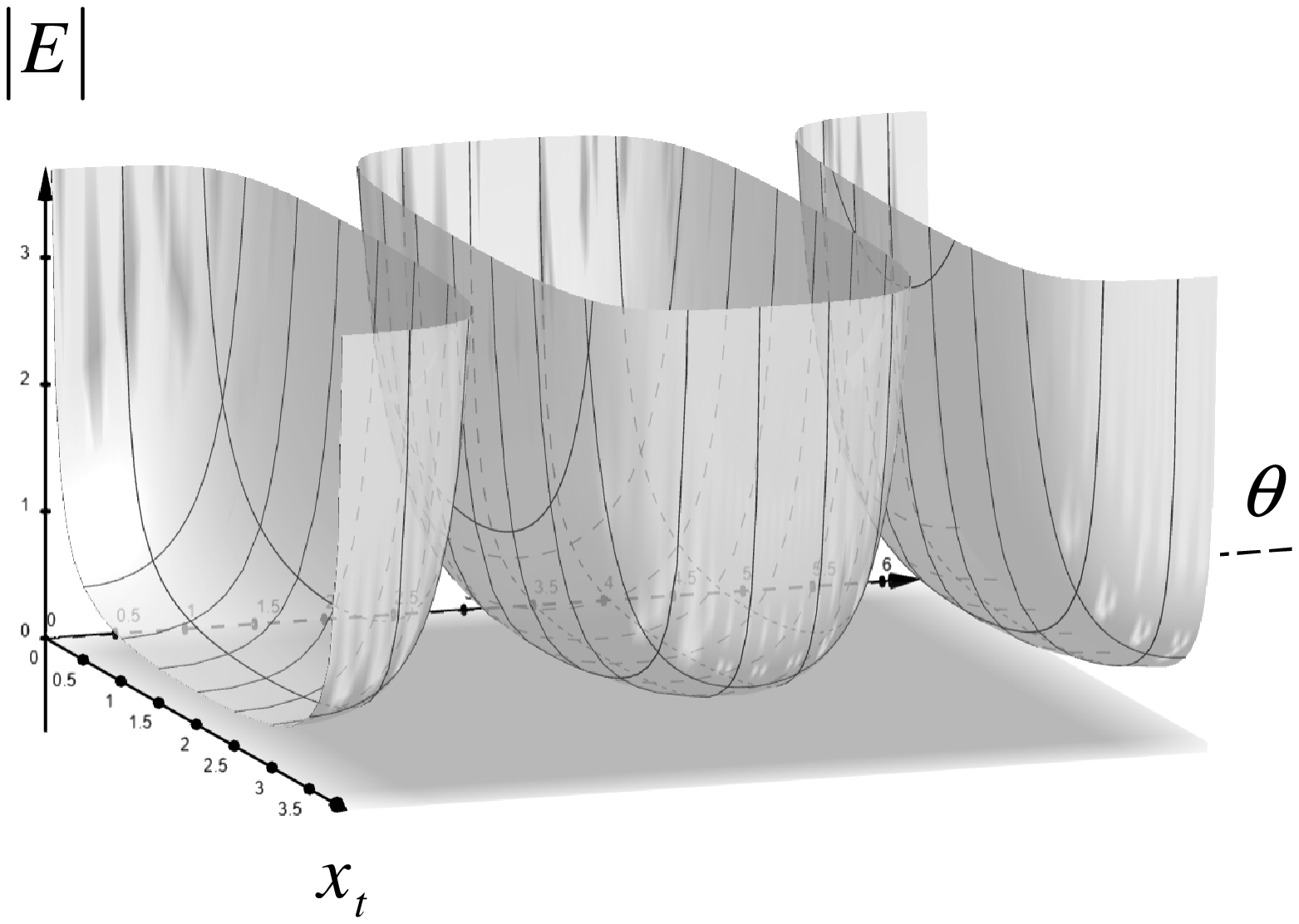}
\caption{\footnotesize 3D graph of $\vert E\vert={k\over (2^2-(x_t-2)^2)\cos^2\theta}$ for k=1, for all $x_t\in[0,4]$, and for all  $\theta\in [0,{2\pi}]\backslash \{{\pi\over2}, {3\pi\over2}\}$. This is the graph of (\ref{CSch}) for $r=2$, $n=0$, $x_t\in[0,4]$ and ${\hbar^2\over 4mr^2}=k=1$. The graph of the energy (\ref{CSch}) in 3D is similar to the graph for a small radius $r$ and a small constant $k={\hbar^2\over 4mr^2}$.
Fluctuation of the energy magnitude  $\vert E\vert$ in the $x_t$-axis and $\theta$-axes. }\label{Fig.9}
\end{minipage}
\hspace*{\fill}
\begin{minipage}[t]{6cm}
\centering
\includegraphics[width=6cm]{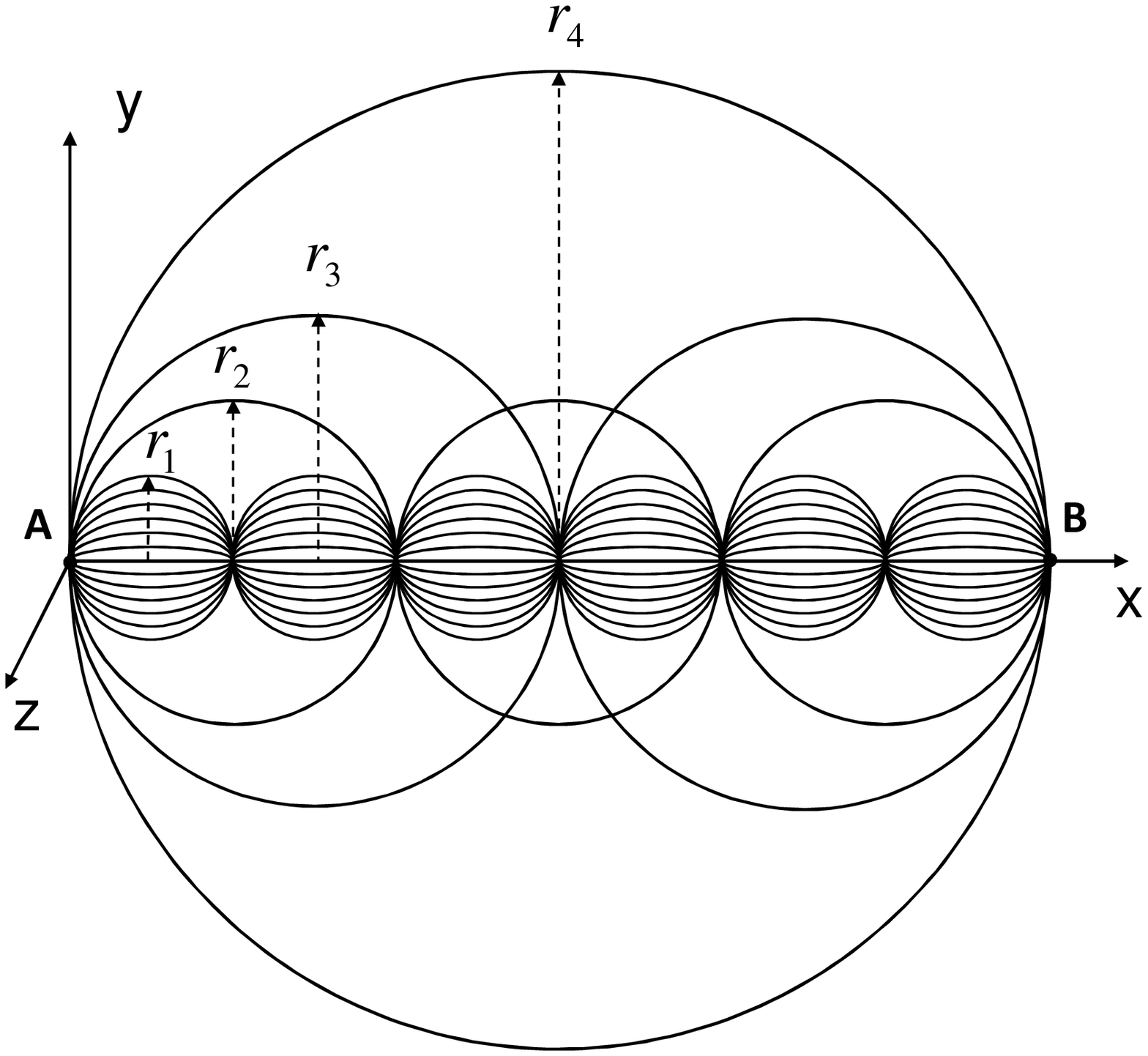}
\caption{\footnotesize Four imbedded families of fluctuating paths of least time between A and B with radius $r_1<r_2<r_3<r_4$. The 4 families represent paths of least time between A and B (with equal length 6$r_1\pi$). If $E_{0_i}$ i=1,..,4 is the zero point energy of the physical system following the geodesics (\ref{Geod3dt}) of radius $r_i$ for i=1,..,4, then $E_{0_1}>E_{0_2}>E_{0_3}>E_{0_4}$ for $r_1<r_2<r_3<r_4$.}\label{Fig.15}
\end{minipage}
\end{figure}

Example of energy variation in 3D with respect to $(x_t,\theta)$ is illustrated in (Fig.\ref{Fig.9}).

\subsubsection{Space minimal energy fluctuation}

The minimum of the energy magnitude (\ref{CSch}) on $]2nr,(2n+2)r[$, for all $\theta\in]0,{2\pi}[\backslash \{{\pi\over2},\pi, {3\pi\over2}\}$ can be found. Indeed, since
\begin{equation}\label{CSch0}
X_{x_t\theta}^2=\Big(r^2-(x_t-(2n+1)r)^2\Big)\cos^2\theta,
 \end{equation}
then $0<X_{x_t\theta}^2\leq r^2\cos^2\theta$, which gives
\begin{equation}\label{CSch1}
\vert E\vert_\theta:={\hbar^2\over4m r^2\cos^2\theta}\leq\vert E\vert,\qquad\forall\theta\in]0,{2\pi}[\backslash \{{\pi\over2},\pi, {3\pi\over2}\},
\end{equation}
the minimum $\vert E\vert_\theta$ is reached at $x_t=(2n+1)r$ (the local fluctuation center of the paths of least time).
The use of all paths of least time by the physical system exhibits a periodic variation of the space minimal energy $\vert E\vert_\theta$ of period $\pi$, see illustration of energy variation  (Fig.\ref{Fig.13}). Indeed,

\textbf{i)} The space minimal energy $\vert E\vert_\theta$ is constant for each path of least time,

\textbf{ii)} The space minimal energy $\vert E\vert_\theta$ is variable for all paths of least time function of $\theta$.

\textbf{iii)} Since energy cannot be created or destroyed, the variation of space minimum energy $\vert E\vert_\theta$ is consequence of a transfer of part of energy to a change of direction in angle $\theta$ (change of path). A change of paths of least time induces a change of the space minimum energy $\vert E\vert_\theta$.

\subsubsection{The zero point energy}

The magnitude $\vert E\vert_\theta$ define the space minimal energy for each path of least time and by continuity of the limit of $\vert E\vert_\theta $ when $\theta$ tends to $0$ or $\pi$ or $2\pi$ with $\theta \neq{0, \pi,2\pi}$ gives
\begin{equation}\label{CSch2}
\lim_{\theta\rightarrow {0, \pi,2\pi}}\vert E\vert_\theta={\hbar^2\over4m r^2}=\vert E\vert_0.
\end{equation}
where $\vert E\vert_0$ is the ground state (the zero point energy) of the physical system for all $(t,\theta)\in [t_a,t_b]\times ]0,{2\pi}[\backslash \{{\pi\over2},\pi, {3\pi\over2}\}$, and it is characterized based on this case study as follow:

\textbf{i)} Since $\vert E\vert_0\neq0$ for all $(t,\theta)\in [t_a,t_b]\times ]0,{2\pi}[\backslash \{{\pi\over2},\pi, {3\pi\over2}\}$, then the ground state energy $\vert E\vert_0$ of the physical system for all paths of least time can never be zero.

\textbf{ii)} The ground state energy $\vert E\vert_0$ for all paths of least time is always constant, and it does not fluctuate. The ground state energy depends on the fluctuation radius of paths of least time and the physical system mass.

\textbf{iii)} The greater the radius of paths of least time (\ref{Geod3dt}) is, the smaller the zero point energy of the physical system is. That is to say, the smaller $\lambda=4r$ (the periodic fluctuation length of each path of least time) the bigger the zero point energy of the physical system is.

\textbf{iv)} The zero point energy can decrease if the physical system changes from an infinity of fluctuating geodesics (\ref{Geod3dt}) with radius $r_1$  to another infinity of fluctuating geodesics (\ref{Geod3dt}) with radius $r_2>r_1$ between two distant locations under the action of an external form of energy (photon for example), then the zero point energy of the physical system will decrease due to this interaction by mean of part of the zero point energy transferred to increase the fluctuation radius of paths of least time. However the lengths of the paths of least time (\ref{Geod3dt}) between the two distant locations with radius $r_1$ or  $r_2$ are equal with conservation of the least time (see Fig.\ref{Fig.15}). The variation of the zero point energy indicates a change of the physical system fluctuation under the effect of external interaction (the photon). This is what happens when observation is used in Young's double-slit experiment.

\subsection{Space minimal energy variation and torque effect interpretation}
Investigation about the Van Der Waals torque between anisotropic parallel plates can be elaborated using this formalism. Indeed
 a torque phenomenon is normally generated in the presence of parallel plates with anisotropic optical properties (such as birefringent crystals) causing a mechanical rotation towards the alinement of the plates principal axes. Interpretation of the Van Der Waals torque can be approached using the fluctuating paths of least time.

\begin{figure}[!h]
\centering
\includegraphics[width=6cm]{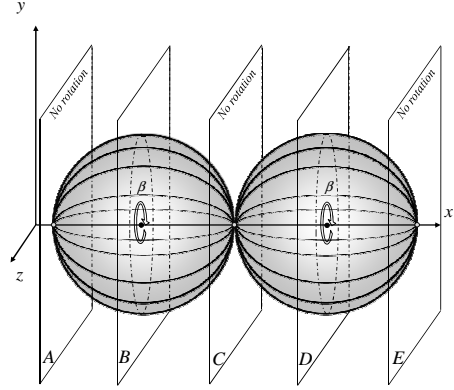}
\caption{\footnotesize Positions where the Van der Waals torque effect can be measured for anisotropic plates. If the plate is positioned on an antipodal point in positions A, C, and E, then there is no torque effect. The Van Der Waals torque can be measured at any position between plates A and C or between plates C and E.}\label{Fig.10}
\end{figure}

When anisotropic plate is placed (with oblique position of the optical axis) in a fixed position $x_t\in ]2nr,(2n+2)r[$ transversal to the paths of least time principal axis direction of the physical system (the $x_t$-axes), the energy (\ref{CSch}) of the physical system in the transversal plan is a convex function of $\theta$ (Fig.\ref{Fig.13}) and constant for each $\theta\in [0,{2\pi}]\backslash \{{\pi\over2}, {3\pi\over2}\}$. The phenomenon that generates the torque is a consequence of the variation of energy on the plate plan for each $\theta\in [0,{2\pi}]\backslash \{{\pi\over2}, {3\pi\over2}\}$ due to the birefringent property of the material plate that creates a circular polarization (see circular polarization \cite{4},\cite{8},\cite{17}). Indeed, when the physical system enters the plate following a path of least time for a given angle $\theta$ in a direction that is oblique with respect to the optical axis of the plate, the birefringent material changes the path of least time of the physical system by double refraction (due to the existence of a variation in refractive index that is sensitive to direction in anisotropic material) which changes the angle $\theta$ of the path of least time in the plate plan from a fluctuating linear motion to helicoidal motion (circular polarization) that generates variation of the energy magnitude (\ref{CSch}) of the physical system in the plate plan.
This variation of energy for each path is transferred into a mechanical rotation. Thus, if the motion of the physical system is blocked, the plate will rotate back to the equilibrium position that defines the stable initial energy at the position $x_t$ in the plan plate.

The use of the infinity of paths of least time (illustrated in Fig.\ref{Fig.00}) in the interpretation of Van der Waals torque effect suggests that:

\textbf{a)} If the plate is positioned at an antipodal point ($x_t=2nr$, or $x_t=(2n+2)r$) then there is no torque effect generated on the plate since there is no possible change of paths of least time on those points (Fig.\ref{Fig.10}).

\textbf{b)} The torque effect can be measured for all positions $x_t\in]2nr,(2n+2)r[$ for all $n\in\nN$. Indeed, the presence of a metal plate optically anisotropic in a transversal position to paths of least time principal axis changes the energy magnitude (\ref{CSch}) of the physical system function of $\theta$, thus the rate of change of (\ref{CSch}) with respect to $\theta$ using (\ref{CSch0}) gives
\begin{equation}\label{CSch3}
F={\partial \vert E\vert\over\partial\theta} =\Big({\hbar^2\over4m X_{x_t\theta}^2\cos^2\theta}\Big){\sin2\theta}=2\vert E\vert\tan\theta
\end{equation}
which represents the amount of rotational energy gained by infinitesimal rotation. The position where this amount of energy is extremum is determined through the solution of equation ${\partial F\over\partial x_t}=0$ that is to say
\begin{equation}\label{CSch4}
{\partial F\over\partial x_t}={\partial^2 \vert E\vert\over\partial x_t\partial\theta} =2{\partial\vert E\vert\over\partial x_t}\tan\theta={\hbar^2\tan\theta\over m\cos^2\theta}{(x_t-(2n+1)r)\over \Big(r^2-(x_t-(2n+1)r)^2\Big)^2}=0
\end{equation}
which gives that the amount of rotational energy gained by infinitesimal rotation is minimal or maximal at the fluctuation center $x_t=(2n+1)r$ according to the sign of $\tan\theta$. Thus

\textbf{i)} For $\tan\theta>0$, the amount of rotational energy gained by infinitesimal rotation is a minimum at the fluctuation center $x_t=(2n+1)r$ and equal to
\begin{equation}\label{CSch5}
F=\Big({\hbar^2\over2m r^2\cos^2\theta}\Big)\tan\theta
\end{equation}

\textbf{ii)} For $\tan\theta<0$, the amount of rotational energy gained by infinitesimal rotation is maximum at the fluctuation center $x_t=(2n+1)r$ and equal to (\ref{CSch5})
\begin{rem}
a) For $\theta\in ]0,{\pi\over2}[\cup ]\pi,{3\pi\over2}[$ we have $\tan\theta>0$, and for $\theta\in ]0,{-\pi\over2}[\cup ]-\pi,{-3\pi\over2}[$ we have $\tan\theta<0$, then the amount of rotational energy gained by infinitesimal rotation at each geodesics local center of fluctuation $x_t=(2n+1)r$ is minimal or maximal based on the rotation direction clockwise or anticlockwise and is given by (\ref{CSch5})

b) Within 1.1 mm there is an average of 2000 light wave lengths that correspond to 4000 aligned spheres of radius r=0.0001375 mm using the geodesics illustrated in (Fig.\ref{Fig.00}) with 4001 antipodal points. Then based on this case study within 1.1 mm there are 4001 positions where the Van der Waals torque is null which explains the difficulty to measure Van der Waals torque effect experimentally.
\end{rem}

\subsection{About the attractive repulsive Casimir force origin}
Investigation about attractive-repulsive Casimir phenomena in vacuum between two distant parallel uncharged plates can be elaborated using this formalism. Indeed the placement of a first uncharged perfectly conducting plate 1 in a position transversal to the paths of least time of the considered physical system induces a change of its energy magnitude (\ref{CSch}). The rate of change of this energy (\ref{CSch}) with respect to paths of least time $X_{x_t\theta}$ is given by
\begin{equation}\label{CSchrodP10}
{\partial \vert E\vert\over\partial X_{x_{t}\theta}} ={-\hbar^2\over2m X_{x_t\theta}^3}.
\end{equation}

Meanwhile, the placement of a second uncharged perfectly conducting plate 2, closed and parallel to the plate 1, in a position transversal to the paths of least time of the considered physical system induces a second changes of its energy magnitude, and its rate of change  with respect to paths of least time $X_{x_t\theta}$ is given by
\begin{equation}\label{CSchrodP11}
{\partial^2\vert E\vert\over\partial X_{x_t\theta}^2} ={3\hbar^2\over2m X_{x_t\theta}^4}.
\end{equation}

This last quantity indicates how fast the magnitude  $\vert E\vert$ is increasing or decreasing with respect to geodesics variation. Thus, equality (\ref{CSchrodP11}) represents an acceleration experienced by the physical system due to the change of energy (consequences of the placement of the two closed plates). This local change of energy (\ref{CSchrodP11}) is transformed into a local force experienced by the physical system given by
\begin{equation}\label{CSchrodP12}
F= m{\partial^2 \vert E\vert\over\partial X_{x_t\theta}^2}={3\hbar^2\over 2X_{x_t\theta}^4},
\end{equation}
this force is also experienced by the plates and might be the origin of Casimir effect. Indeed, the use in $X_{x_t\theta}$ of the parameter
\begin{equation}\label{C1}
x_t=\pm\sqrt{r^2-{D^2\over\cos^2\theta}\sqrt{{360\hbar\over c\pi^2}}}+(2n+1)r
\end{equation}
for all $\theta\in]0,{2\pi}[\backslash \{{\pi\over2},\pi, {3\pi\over2}\}$ gives
\begin{equation}\label{C2}
X^2_{x_t\theta}=\Big(r^2-(x_t-(2n+1)r)^2\Big)\cos^2\theta=D^2\sqrt{{360\hbar\over c\pi^2}}
\end{equation}
which gives in (\ref{CSchrodP12})
\begin{equation}\label{CSchrodP13}
F={3\hbar^2\over2 X_{x_t\theta}^4}={3\hbar^2 c\pi^2\over2 {D^4 360\hbar}}={\hbar c\pi^2\over240 {D^4}},
\end{equation}
which is the magnitude of Casimir force (\cite{5}) felt by the plates by unity of surface, where c is the speed of light , $\hbar$ is the reduced Planck constant, and D is the distance that separates the plates.

\subsection{Attractive or repulsive Casimir force causality interpretation}

\begin{figure}[!h]
\begin{minipage}[t]{6cm}
\centering
\includegraphics[width=6cm]{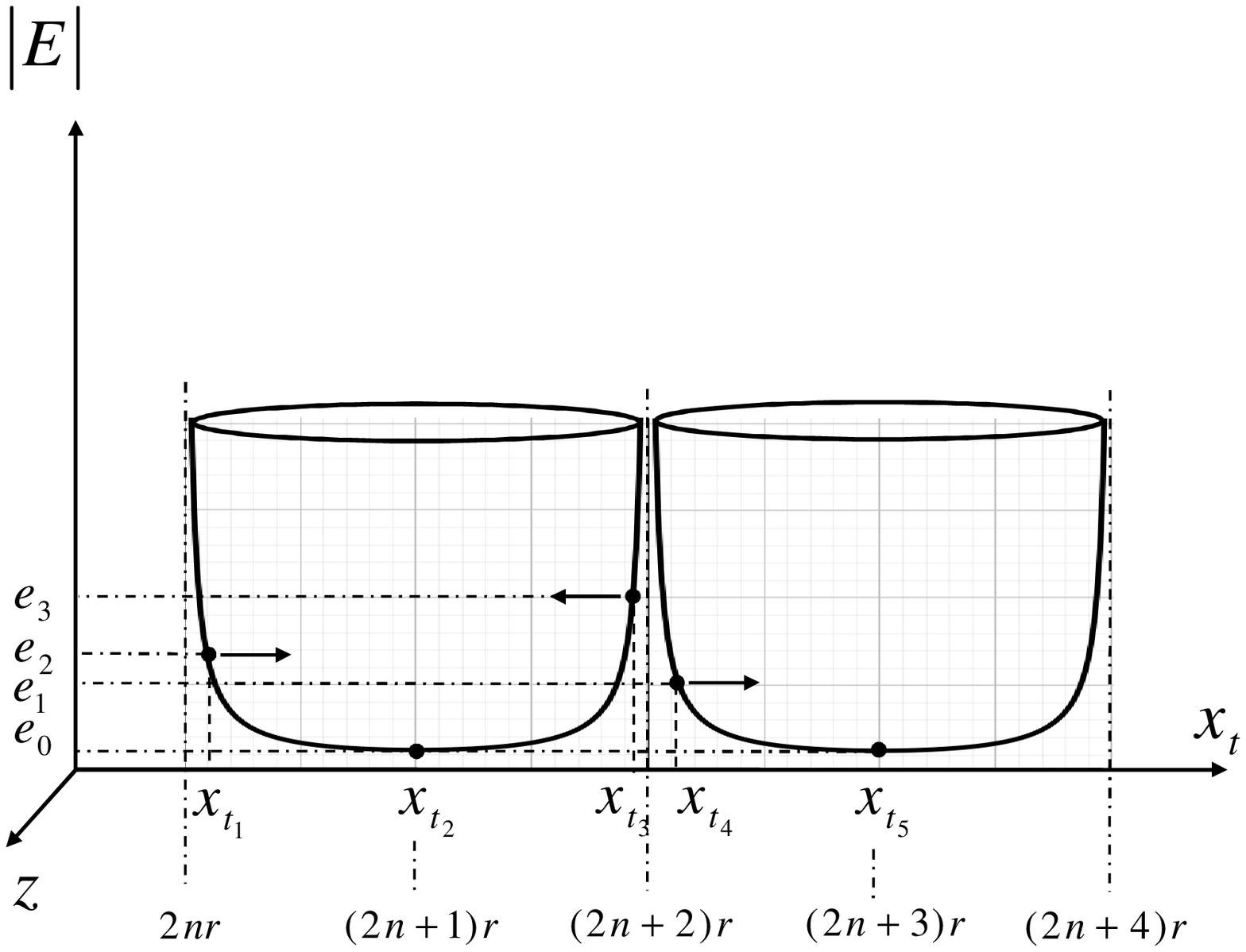}
\caption{\footnotesize Illustration of position and energy $\vert E\vert$ variation given by (\ref{CSch}) on $]2nr,(2n+2)r[\times(]0,{2\pi}[\backslash \{{\pi\over2},\pi, {3\pi\over2}\})$ for $n$ integer. The presence of plates A at $x_{t_1}$ and  C at $x_{t_3}$ induces energy variation, a transfer of part of energy into a mechanical force (\ref{CSchrodP13}) experienced by the plates A and C to move to a position with lower energy (after transfer). The lower energy position is located in the direction of each arrow. No attraction/repulsion at the positions $x_{t_2}$ and $x_{t_5}$}\label{Fig.11}
\end{minipage}
\hspace*{\fill}
\begin{minipage}[t]{6cm}
\centering
\includegraphics[width=6cm]{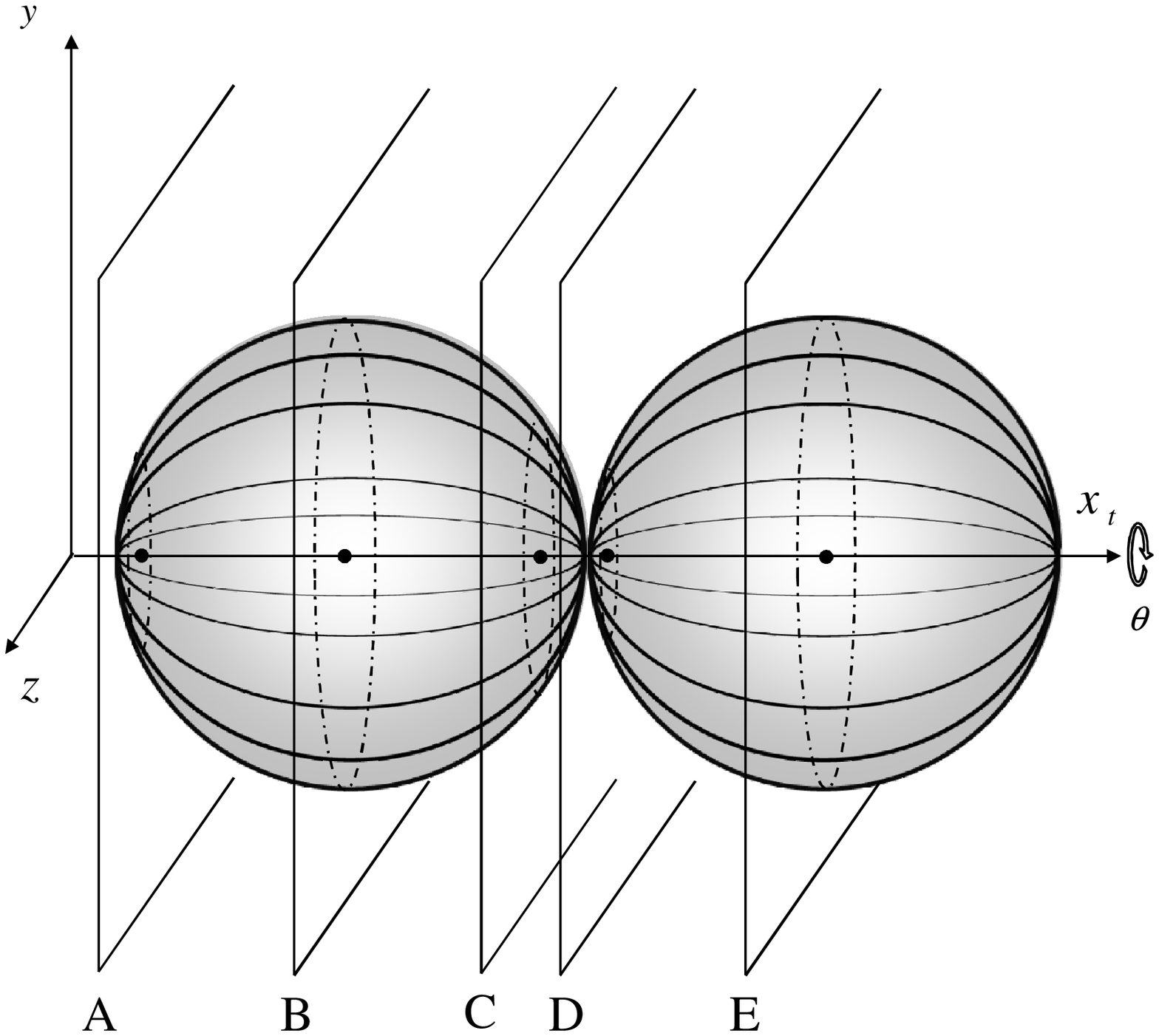}
\caption{\footnotesize Parallel plates: Based on illustration (Fig.\ref{Fig.11}) the plate A cross the $x_t$-axis at $x_{t_1}$, the plate B cross the $x_t$-axis at $x_{t}=(2n+1)r$ , the plate C cross the $x_t$-axis at $x_{t_3}$ , the plate D cross the $x_t$-axis at $x_{t_4}$, the plate E cross the $x_t$-axis at $x_{t}=(2n+3)r$. All plates are parallel to the yz-plan. Only two plates are considered for any interpretation of the attraction or repulsion of the plates under Casimir force. }\label{Fig.12}
\end{minipage}
\end{figure}

The magnitude energy $\vert E\vert$ defined in (\ref{CSch}) of the physical system is a convex function on the product $]2nr,(2n+2)r[\times\Big(]0,{2\pi}[\backslash \{{\pi\over2},\pi, {3\pi\over2}\}\Big)$ for $n\in\nN$ as illustrated in Fig.\ref{Fig.11}.

The magnitude energy (\ref{CSch}) is determined at each $x_t\in]2nr,(2n+2)r[$, and when a plate A is introduced in a position $x_{t_1}$ (Fig.\ref{Fig.11}) parallel to the $\vert E\vert z$-plan located at the left side of the fluctuation center $x_t=(2n+1)r$ followed by the introduction of a second parallel plate C in the the position $x_{t_3}$ at the right side of the fluctuation center $x_t=(2n+1)r$ on paths of least time (Fig.\ref{Fig.11} and Fig.\ref{Fig.12}), perturbation of the energy state of the physical system is created and given by (\ref{CSchrodP11}).
Since energy cannot be created or destroyed, rather it can only be converted into another form of energy then the perturbation of the energy is manifested by a transfer of part of the physical system energy into a mechanical force (\ref{CSchrodP13}) experienced by the plates to move from positions $x_{t_1}$ and $x_{t_3}$ to a new position with lower energy after transfer (see Fig.\ref{Fig.11}).
The lower energy position of the plate A will be located to its right and a lower energy position for the plate C will be located to its left based on the graph of the energy magnitude illustrated in (Fig.\ref{Fig.11}). The two plates will experience the Casimir attractive force.
However, if the considered plates are the plate C and the plate D at the position $x_{t_3}$ and $x_{t_4}$ (Fig.\ref{Fig.11}, and Fig.\ref{Fig.12}), the lower energy position for the plate C will be located to its left and the lower energy position for the plate D will be located to its right based on the graph of the energy magnitude. Therefore the two plates will experience the Casimir repulsive force.
The repulsive force or attractive force are functions of the position of plates with respect to the fluctuation centers of paths of least time  $x_t=(2n+1)r$ in each local interval $]2nr,(2n+2)r[$ for all $n\in\nN$ and we have:

\textbf{i)} The repulsive or attractive Casimir force is due to the energy magnitude variation of the physical system, consequence of the presence of closed plates. This variation occurs by mean of a transfer of part of energy of (\ref{CSch}) at $x_t$ to repulsive or attractive motion. When a part of energy (\ref{CSch}) at $x_t$ is transferred, the plate will move to a position $x_{t'}$ that fits the remaining energy (lower energy). The plates under Casimir force always move to a new position with lower energy after transfer. Nevertheless, the effect of Casimir force is local because the perturbation of the energy is local.

\textbf{ii)} Attractive Casimir force: if the two positions are located at the left and right side of the local fluctuation center $x_t=(2n+1)r$ then the two plates will experience attraction. Attraction is also experienced for two plates when one is positioned in the left side of a local fluctuation geodesic center, and a second positioned in the right side of the next local fluctuation geodesic center (Fig.\ref{Fig.11}).

\textbf{iii)} Repulsive Casimir force: if one plate is located at the right side of the center of the local interval $]2nr,(2n+2)r[$ and a second plate is located at the left side of the center of the local interval $](2n+2)r,(2n+4)r[$ or $](2n+4)r,(2n+6)r[$,  then the two plates will experience a repulsive force (Fig.\ref{Fig.11}).

\textbf{vi)} if the two positions are located at two centers of fluctuation $x_{t_1}=(2n+1)r$ and $x_{t_2}=(2n+3)r$ for all $n\in \nN$, then the plates won't experience neither attraction nor repulsion. This case can be justified. Indeed, we have ${\partial \vert E\vert\over\partial X_{x_{t}\theta}}\Big\vert_{x_t=(2n+1)r}={\partial \vert E\vert\over\partial x_{t}}{\partial x_t\over\partial X_{x_{t}\theta}}=0$ since ${\partial \vert E\vert\over\partial x_{t}}={\hbar^2(x_t-(2n+1)r)\over2m\cos^2\theta(r^2-(x_t-(2n+1)r)^2)^2}\Big\vert_{x_t=(2n+1)r}=0$. Moreover at these positions the energy is minimal and it cannot be reduced to a lower energy level, then there is no part of the energy that can be transferred into another form (Fig.\ref{Fig.11}).

\section{Conclusion}

The use of an infinity of fluctuating paths of least time that are compatible with the quantum mechanics indeterminacy provides a new interpretation in geometrical optic of the interference pattern of Young's double-slit experiment, not only in the detector screen but also in the whole geodesics intersection region between the slits and detector screen, which suggests that the wave aspect of matter and radiation (photons, electrons, neutrons, atoms and molecules) is actually dictated by the space-time local geometry. Matter wave nature is merely arising from interaction of the physical systems with the space-time geodesics that induces a periodic fluctuating energy.
Following the essence of Feynman path integral, the association of a density of probability of presence of the physical system to each used path of least time (as well as a wave function) leads to derive the Schr\"{o}dinger equation starting from the geodesic's characteristics together with an uncertainty relation between position an momentum.
The energy of the physical system extracted from the Schr\"{o}dinger equation has provided a valuable information about the origin of the phenomenon of the Van Der Waals torque between anisotropic parallel plates, as well as a clear mechanism of the attractive/repulsive Casimir force.
Energy variation appears to be a manifestation of the interaction of the physical system with the space time geodesics, where the ground state energy is a residual energy (the minimal energy limit) that cannot vanish. It is the minimum energy conceived by the physical system that characterizes the fluctuation radius of paths of least time in the medium. This is how the space-time dictates the behavior of the physical system at the quantum scale based on this case study.
%\vskip6pt

\enlargethispage{20pt}

%\ethics{Insert ethics text here.}

%\dataccess{This article has no additional data.}

%\aucontribute{Insert author contributions text here (to be included if more than one author).}

%\competing{Insert competing interests text here.}

%\funding{Insert funding text here.}

%\ack{Insert acknowledgment text here.}

%\disclaimer{Insert disclaimer text here.}

%%%%%%%%%% Insert bibliography here %%%%%%%%%%%%%%

\end{document}